\documentclass[12pt]{article}
\usepackage[utf8]{inputenc}
\usepackage{booktabs}
\usepackage{cite}
\usepackage{amsmath,amssymb,amsfonts}
\usepackage{amssymb}
\usepackage{algorithmic}
\usepackage{graphicx}
\usepackage{textcomp}
\usepackage{xcolor}
\usepackage{xspace}
\usepackage{color}
\usepackage{tcolorbox}
\usepackage{multirow}
\usepackage{multicol}
\usepackage{booktabs}
\usepackage[caption=false]{subfig}
\usepackage{url}
\usepackage{balance}
\usepackage{xspace}
\usepackage{float}
\usepackage{setspace}
\usepackage[paperwidth=8.5in,paperheight=11in,left=1in,right=1in,top=1.5in,bottom=1.5in]{geometry}
\usepackage{picins}

\usepackage{algorithmic}
\usepackage[ruled, linesnumbered]{algorithm2e}
\usepackage{algorithmic,algorithm2e,float}

\newcommand{\mcode}[1]{$\tt #1$}

\usepackage{array}
\usepackage{ragged2e}
\newcolumntype{P}[1]{>{\RaggedRight\hspace{0pt}}p{#1}}

\begin{document}
\onehalfspacing

\begin{center}
{\Large \textbf{Towards a Catalog of Composite Refactorings}}\\

\vspace{6mm}
Aline Brito, Andre Hora, Marco Tulio Valente\\
\vspace{4mm}

{\small
{Department of Computer Science,\\ Federal University of Minas Gerais (UFMG), Brazil}\\
\{alinebrito, andrehora, mtov\}@dcc.ufmg.br
}

\vspace{6mm}

\parbox{0.85\textwidth}{\noindent\textbf{Abstract.}
Catalogs of refactoring have key importance in software maintenance and evolution, since developers rely on such documents to understand and perform refactoring operations. Furthermore, these catalogs constitute a reference guide for communication between practitioners since they standardize a common refactoring vocabulary. Fowler's book describes the most popular catalog of refactorings, which documents single and well-known refactoring operations. However, sometimes refactorings are composite transformations, i.e., a sequence of refactorings is performed over a given program element. For example, a sequence of Extract Method operations (a single refactoring) can be performed over the same method, in one or in multiple commits, to simplify its implementation, therefore, leading to a Method Decomposition operation (a composite refactoring). In this paper, we propose and document a catalog with eight composite refactorings. We also implement a set of scripts to mine composite refactorings  by preprocessing the results of  refactoring detection tools. Using such scripts, we search for composites in a representative refactoring oracle with hundreds of confirmed single refactoring operations. Next, to complement this first study, we also search for composites in the full history of  ten well-known open-source projects. We characterize the detected composite refactorings, under dimensions such as size and location. 
We conclude by addressing the applications and implications of the proposed catalog. }
\end{center}


\section{Introduction}
\label{section:introduction}

Refactoring is a fundamental practice to keep software in a healthy shape. Developers have learned over the years that the lack of continuous refactoring can rapidly transform software projects into a ``big-ball-of-mud''~\cite{Modularization:Sarkar:2009,BigBallofMud:1997}. As a result, maintenance---including both bug fixes and the implementation of new features---becomes very risky and slow. 
Therefore, to promote and facilitate the dissemination of this practice among developers, refactoring operations are usually documented in catalogs, like the one proposed by Fowler in his seminal book published in 1999~\cite{Fowler:1999}. In this catalog, Fowler provides detailed documentation about dozens of refactorings, providing a name for each refactoring, describing the mechanics required to perform the source code transformation, and also giving illustrative examples of the proposed refactorings.
However, most refactorings described in Fowler’s catalog are restricted in time and scope. 
Particularly, they are described as source code transformations that can be performed by a single developer, in a short time frame (time constraint) and by impacting a limited number of program elements (scope constraint). 
This understanding of refactoring is also assumed by modern refactoring detection tools, such as Refactoring Miner~\cite{Tsantalis:2020:RefactoringMiner2,Tsantalis:ICSE:2018:RefactoringMiner,pyref:scam:2021} and RefDiff~\cite{danilo:tse2020:refdiff2,danilo:msr2017:RefDiff,Brito:2020:RefDiff4Go}.  
Indeed, these tools report refactorings at a very fine granularity level. 
For example, suppose that a given method \mcode{m()} is implemented in classes \mcode{A1}, \mcode{A2},…, \mcode{An}. 
Then, suppose that a Pull Up refactoring is performed to move the replicated method to a superclass B.
These tools report this refactoring as a sequence of the following unrelated operations:

\begin{verbatim}
Pull up: A1.m() to B
Pull up: A2.m() to B
Pull up: A3.m() to B
...
\end{verbatim}

However, we claim that the best output would be reporting a single composite refactoring operation:

\begin{verbatim}
Pull up: A1.m(), A2.m(), A3.m(), …, to B
\end{verbatim}

This is just a trivial example of composite refactoring (in Section \ref{section:catalog} we provide a more complex example). 
Indeed, composite refactorings were previously defined by Souza et al. as \textit{``two or more interrelated refactorings that affect one
or more elements''}~\cite{Sousa:2020:Composite}. However, in their work, the authors focused on the role played by composite refactorings when removing code smells. 
In other words, they do not explore,  document, and illustrate a comprehensive catalog of composite refactorings, which is exactly our goal in this paper.

We initially describe eight composite refactorings, in abstract terms and using illustrative examples. Then, we implemented a set of scripts to identify these composite refactorings. To conclude, we used these scripts to mine composite refactorings in two datasets, as follows:

\medskip

\noindent{\em Oracle study:} 
In the first study, we mine composites in a large and representative refactoring oracle commonly used in the literature~\cite{Tsantalis:2020:RefactoringMiner2,TsantalisOracle}. Specifically, we look for occurrences of the refactorings in our catalog among 1.7K confirmed single refactoring instances listed in this oracle.  We identify that {\bf a significant rate of 60\% of the refactorings of  interest in this oracle are part of composite operations}.  We also characterize the detected composite refactorings, under dimensions such as size and scope.

\medskip

\noindent{\em Study in the wild:} In the oracle study, we rely on a  sample that includes selected refactorings from distinct projects. Therefore, as a complementary analysis, we also look for composite refactorings in the full history of ten popular GitHub projects. As a result, {\bf we were able to identify and characterize 2,886 instances of composite refactorings}.

\medskip

Therefore, our contributions are threefold: (1) we propose a comprehensive catalog of composite refactorings; (2) we implemented a set of scripts to detect the refactorings proposed in this catalog; and (3)
we characterize a large sample of composite refactorings performed in real software projects, including a new viewpoint of a well-known refactoring oracle.\\[-0.2cm]

\noindent{\em Paper Structure:} 
Section \ref{section:firstExample} shows a first example of composite refactoring, while Section \ref{section:catalog} introduces the proposed catalog. 
We describe the results of the oracle study in Section \ref{section:composite-oracle}, while Section \ref{section:composite-wild} includes results in the wild, covering the full evolution history of ten popular open-source projects. 
The results are then discussed in  Section \ref{section:discussion}.  
Section \ref{section:threatsValidity} states threats to validity and 
Section \ref{section:relatedWork} presents related work. 
Finally, we conclude the paper in Section \ref{section:conclusion}.

\section{An Example of Composite Refactoring}
\label{section:firstExample}

\begin{figure}[!th]
	\centering
    \includegraphics[width=1\textwidth]{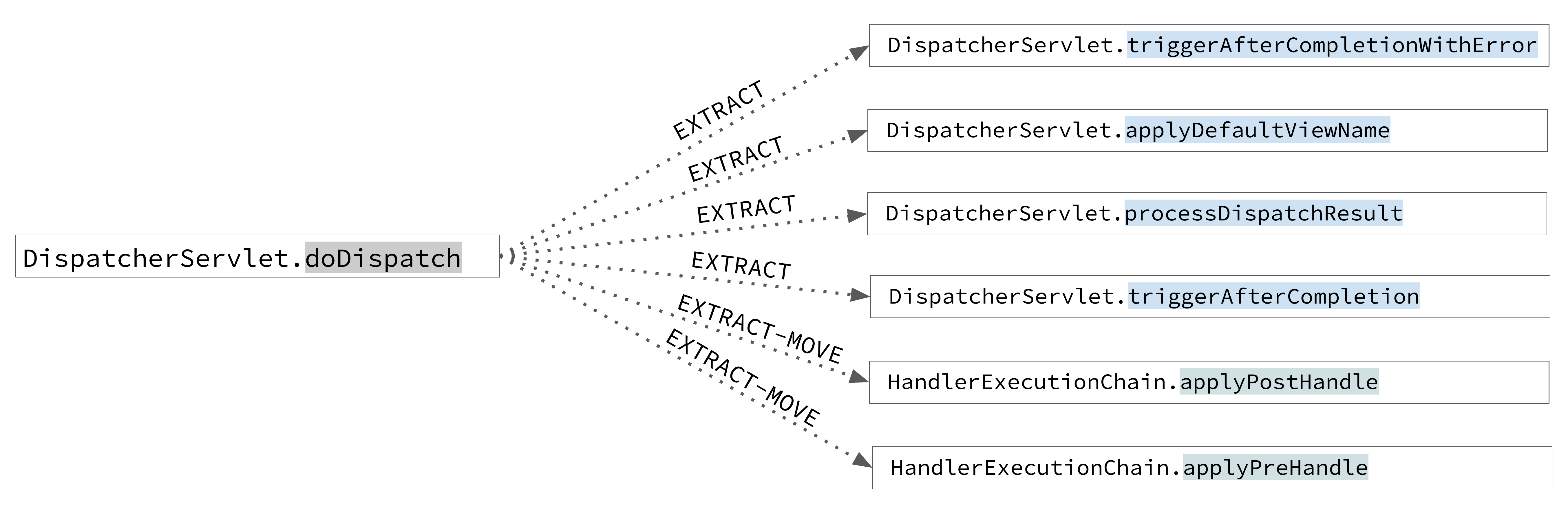}
	\caption{Example of composite refactoring from \textit{Spring Framework}. Method \textit{doDispatch} was decomposed by applying six Extract Method refactorings}
	\label{fig:fig_first_example}
\end{figure}

Figure \ref{fig:fig_first_example} shows a real example of composite refactoring in {\sffamily \small Spring}, a well-known Web framework for Java. 
As we can see, method \mcode{
doDispatch(...)} was decomposed by performing six {\sffamily  \small Extract Method} refactorings. 
Moreover, in two cases the {\sffamily  \small Extract} was followed by a {\sffamily \small  Move Method}. 
These operations were performed in a single commit.\footnote{{github.com/spring-projects/spring-framework/commit/3642b0f3}}

When used in a context like this one, a refactoring detection tool, such as RefDiff~\cite{danilo:tse2020:refdiff2,danilo:msr2017:RefDiff,Brito:2020:RefDiff4Go} or RefactoringMiner~\cite{Tsantalis:2020:RefactoringMiner2,
Tsantalis:ICSE:2018:RefactoringMiner}, detects these six
single refactorings independently. 
However, it would be interesting to detect a high-level refactoring operation, i.e., a composite refactoring grouping the six transformations. 
As we detailed in the next section, we call {\sffamily \small  Method Decomposition} refactoring, this particular coarse-grained operation.

Techniques that may benefit from the detection of independent refactorings (like code visualization~\cite{AlOmar:book:BigDataRefactoring:2021,Merino:JSS:2018,Brito:2020:RefGraph}, code review~\cite{Paixao:2020,bacchelli:2013:ICSE,bacchelli:2018:ICSESEIP,bacchelli2013expectations,ge2017refactoring}, code authorship~\cite{Avelino:icpc:2016,hattori09}, bug-introducing detection~\cite{rahman11a,Kim06a}, refactoring-aware tools~\cite{Shen:OOPSLA:2019,Brito:RAID:icpc:2021}, software mining approaches~\cite{Higo:2020:JSS,Grund:2021:ICSE,hora:icse2018:UntrackedChanges,Spadini:2018:FSE}, to name a few) may also benefit from the detection of composite refactorings.
As refactoring detection is the basis of such techniques, composite refactorings would bring to light novel operations not restricted to time and scope, therefore, better representing the actual source code changes.

Before presenting our catalog, it is important to mention that composite refactorings are not limited to a single commit~\cite{Sousa:2020:Composite}. 
For example, as stated by Fowler in the new version of his book on refactoring~\cite{Fowler:2018}, there are also  \textit{long-term} refactorings  ``\textit{that can take a team weeks to complete}''.

\section{Catalog of Composite Refactorings}
\label{section:catalog} 

In this section, we introduce the proposed catalog of composite refactorings. As customary in refactoring catalogs, we describe the proposed refactoring types and their mechanics.  We also present an abstract example of each composite refactoring. There are two main groups of refactorings: (i) to decompose program structures (two composite refactorings), and (ii) to create program structures (four composite refactorings).
To propose these composites, we basically leveraged our previous research on large refactorings operations. For example, we started this research by characterizing refactoring operations performed over time in Java projects~\cite{Brito:2020:RefGraph}. Later, we extended this work by including JavaScript projects and also performed a survey with the developers who were responsible for these operations~\cite{Brito:2021:RefGraph:EMSE}.

\subsection{Class Decomposition}

\noindent {\em Motivation:} According to Fowler~\cite{Fowler:1999,Fowler:2018}, during software evolution we might need to ``move elements around'', aiming to improve modularity and cohesion and reduce coupling. Specifically, single {\sffamily \small  Move Method} operations should be performed ``when classes have too much behavior or when classes are collaborating too much and are too highly coupled''. However, the overall solution may not be restricted to a single refactoring operation. Instead, we might need to move more than one method from a single source class. In this case, we say we performed a composite refactoring called {\sffamily \small  Class Decomposition}.

\bigskip

\begin{figure}[!th]
	\centering
    \includegraphics[width=0.6\textwidth]{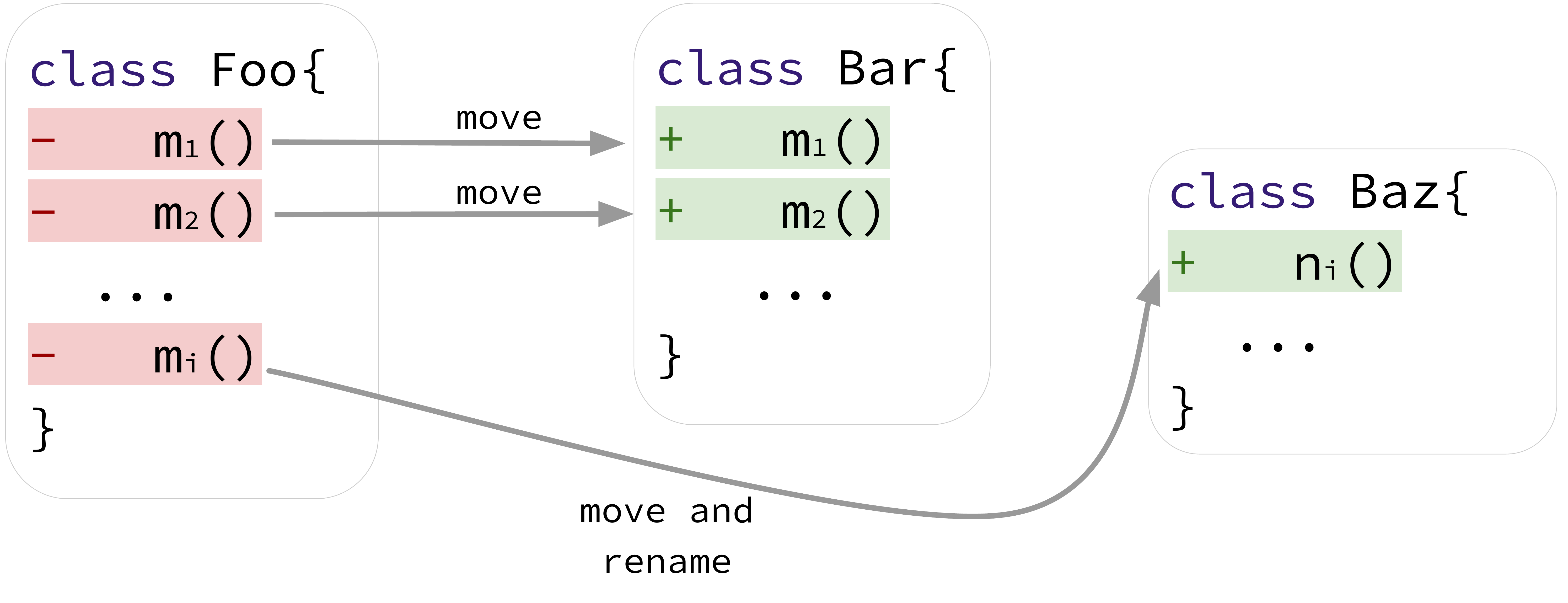}
	\caption{Class Decomposition}
	\label{fig:fig_class_decomposition}
\end{figure}

\noindent {\em Mechanics:} Figure \ref{fig:fig_class_decomposition} shows an example, in which class \mcode{Foo} lost multiple methods to classes \mcode{Bar} and \mcode{Baz}. The target class can be existing or new. Also,  the {\sffamily \small Move} operations can be followed by a {\sffamily \small Rename} operation.  In all cases, the final goal is to decompose the source class and make its implementation more cohesive. It is also worth noting that our definition does not require all move operations to be performed in a single commit. In other words, they can be spread over time, in multiple commits.

\subsection{Method Decomposition}

\noindent {\em Motivation:} We  perform {\sffamily \small Extract Method} operations when  ``you have to spend effort looking at a fragment of code and figuring out what it's doing''~\cite{Fowler:1999,Fowler:2018}. In other words, Fowler advocates the improvement of  understandability as the main reason to perform method extractions.
However, the solution does not need to be limited to a sole operation. We could perform a sequence of two or more {\sffamily \small Extract Method} operations over a single method. As a result, it generates a simpler one. Evidently, these refactorings also generate new methods. However, the goal is still the decomposition of the source method. In this case, we say we performed a composite refactoring called  {\sffamily \small Method Decomposition}.

\bigskip

\begin{figure}[!th]
	\centering
    \includegraphics[width=0.55\textwidth]{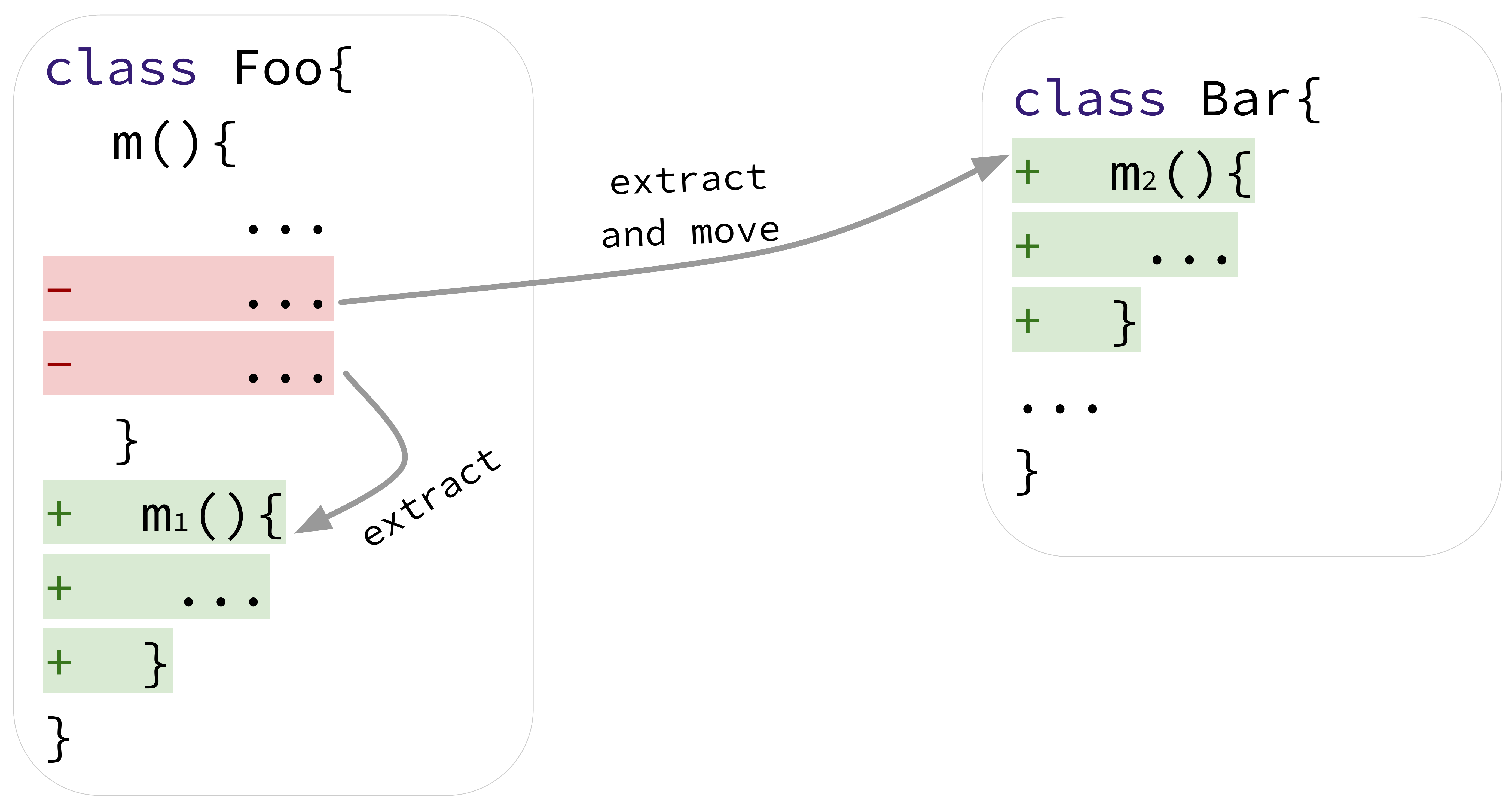}
	\caption{Method Decomposition}
	\label{fig:fig_method_decomposition}
\end{figure}

\noindent {\em Mechanics:} Figure  \ref{fig:fig_method_decomposition} shows an abstract example, in which  methods \mcode{m_1()} and \mcode{m_2()} were extracted from method \mcode{m()}. After the extractions, the new methods can be moved to a distinct class, as happened with \mcode{m_2()}. As usual, the operations can be performed in one or multiple commits.

\subsection{Method Composition}

\noindent {\em Motivation:} Extractions also can be performed to promote reuse and to remove duplication~\cite{Fowler:2018,Fowler:1999,danilo:fse2016:WhyWeRefactor}. Particularly, in such cases, we have similar fragments of code scattered over multiple locations. Therefore, a single {\sffamily \small Extract Method} operation does not eliminate the duplicated issue. Instead, it may be necessary to apply multiple extractions to remove the duplicated code, generating a new method. In this case, we say we performed  a composite refactoring called {\sffamily \small Method Composition}.

\bigskip

\begin{figure}[!th]
	\centering
    \includegraphics[width=0.55\textwidth]{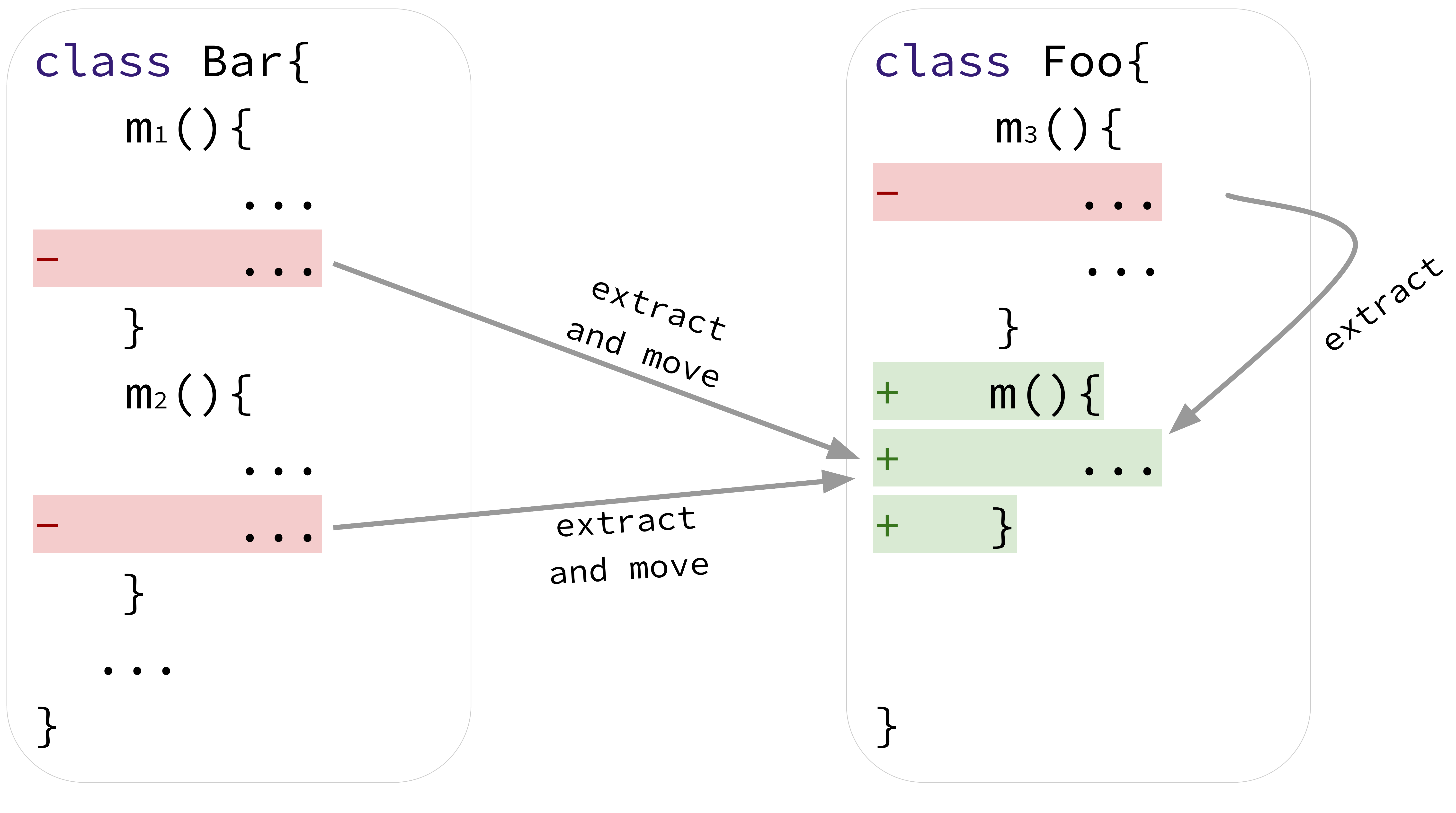}
	\caption{Method Composition}
	\label{fig:fig_method_composition}
\end{figure}

\noindent {\em Mechanics:} Two or more {\sffamily \small  Extract Method} operations are performed over duplicated code, as illustrated in Figure \ref{fig:fig_method_composition}. This code is then removed and a new method is created, with the previously duplicated code. 
The operations also can be followed by {\sffamily \small Move Method} operations, i.e., the new method is placed in a distinct class.

\subsection{Composite Inline Method}

\noindent {\em Motivation:}  {\sffamily \small Inline Method}---as originally proposed in Fowler's catalog---is reported as the opposite operation of {\sffamily \small Extract Method}. 
The author suggests applying a set of  {\sffamily \small Inline Method} operations to remove trivial methods~\cite{Fowler:2018}.
However, {\sffamily \small Inline Method} is usually detected as a single operation by current refactoring detection tools~\cite{danilo:tse2020:refdiff2,Tsantalis:2020:RefactoringMiner2}. That is, such tools report independent {\sffamily \small Inline} operations, even when they are part of the same group of operations.
Therefore, we decided to include this refactoring in our catalog, since it matches our criteria for composite refactorings and is not properly explored and detected by current tools.

\bigskip

\begin{figure}[!th]
	\centering
	\vspace{0.2cm}
    \includegraphics[width=0.5\textwidth]{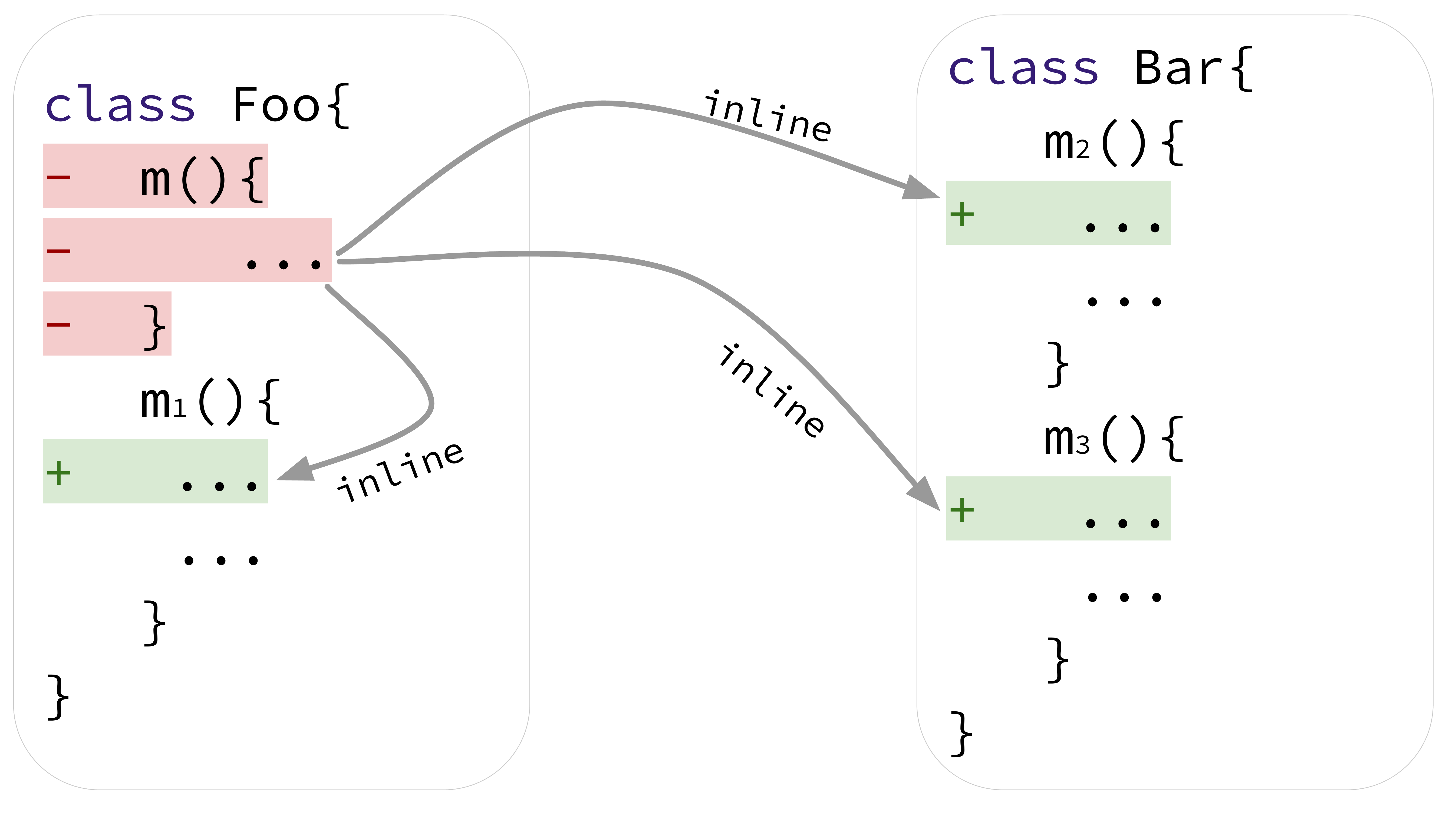}
	\caption{Composite Inline Method}
	\label{fig:fig_inline_decomposition}
	\vspace{0.2cm}
\end{figure}

\noindent {\em Mechanics:}  We expand a (simple) method body in its call sites, as shown in Figure \ref{fig:fig_inline_decomposition}. Then, we remove the source method. The calls may be located in methods from distinct classes.

\subsection{Composite Pull Up Method}

\noindent {\em Motivation:}  Fowler also points to the need to move up or down methods in inheritance hierarchies~\cite{Fowler:2018}. In this context, we apply sequences of {\sffamily \small Pull Up Method} to create a single and more general method in the superclass, therefore achieving code reuse.
As in the case of {\sffamily \small Inline}, we decided to include this refactoring in our catalog mainly because 
{\sffamily \small Pull Up} operations are reported as individual and independent operations by current refactoring detection tools.

\bigskip

\begin{figure}[!th]
	\centering
    \includegraphics[width=0.8\textwidth]{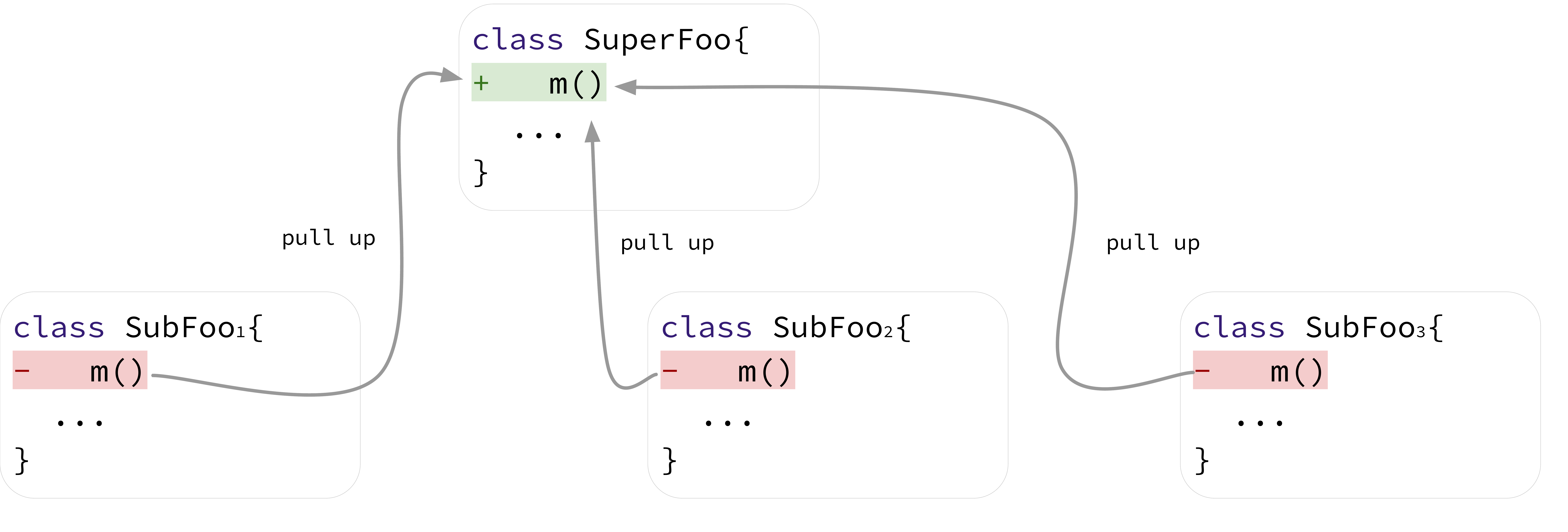}
	\caption{Composite Pull Up Method}
	\label{fig:fig_pull_up_decomposition}
\end{figure}

\noindent {\em Mechanics:} This operation refers to sequences of transformations performed to move methods from subclasses to their superclass.
For example, consider a class \mcode{SuperFoo} with subclasses \mcode{SubFoo1}, \mcode{SubFoo2}, and \mcode{SubFoo3}, as presented in Figure \ref{fig:fig_pull_up_decomposition}. Suppose that a {\sffamily \small Pull Up} operation is applied to move method \mcode{m()} from these subclasses to the superclass. Usually, this operation occurs in a single commit. First, a developer copies the method \mcode{m()} to the superclass, which can be an existing or new one. After that, the method is removed from the subclasses. In this context, the following three messages are issued by RefactoringMiner~\cite{Tsantalis:2020:RefactoringMiner2}:

\medskip

{\small 
\begin{verbatim}Pull Up Method public m() : void from class SubFoo1
to public m() : void from class SuperFoo
\end{verbatim}
}

\smallskip

{\small 
\begin{verbatim}Pull Up Method public m() : void from class SubFoo2
to public m() : void from class SuperFoo
\end{verbatim}
}

\smallskip

{\small 
\begin{verbatim}Pull Up Method public m() : void from class SubFoo3
to public m() : void from class SuperFoo
\end{verbatim}
}

\smallskip

However, since essentially they are part of the same composite refactoring, we claim these operations should have been reported using a single and comprehensive message, such as:

\medskip

{\small 
\begin{verbatim}Pull Up method public m() : void 
From: SubFoo1, SubFoo2, and SubFoo3 
To: public m() : void in SuperFoo
\end{verbatim}
}

\subsection{Composite Push Down Method}

\noindent {\em Motivation:} 
As an opposite scenario, we perform {\sffamily \small Push Down Method} when a method is needed only in a few subclasses~\cite{Fowler:1999,Fowler:2018}. Therefore, this refactoring promotes inheritance simplification. This operation---also present in Fowler's catalog---matches our criteria for composite refactoring. However, as in the case of {\sffamily \small Pull Up} and {\sffamily \small Inline}, it is reported as independent operations by current refactoring mining tools.

\bigskip

\begin{figure}[!th]
	\centering
    \includegraphics[width=0.55\textwidth]{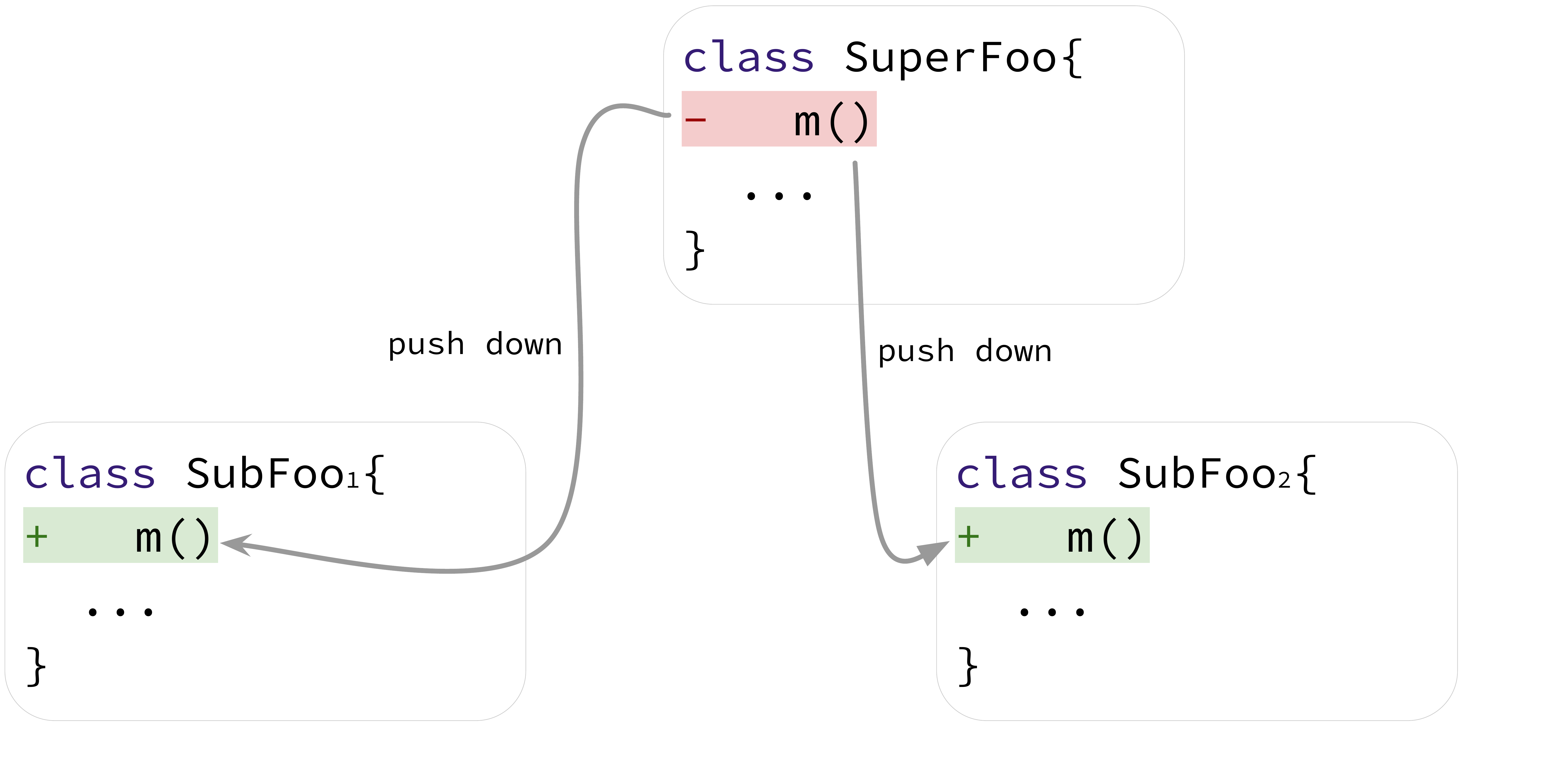}
	\caption{Composite Push Down Method}
	\label{fig:fig_push_down_decomposition}
\end{figure}

\noindent {\em Mechanics:} This operation moves a given method from the superclass to particular subclasses, as presented in Figure \ref{fig:fig_push_down_decomposition}. After that, the method is removed from the superclass.

\subsection{Composite Pull Up Field}

\noindent {\em Motivation:} Often, we have duplicate data in inheritance hierarchies, for example, fields used for a similar purpose in distinct subclasses. 
In this case, we can perform a sequence of {\sffamily \small Pull Up Field} to create a single one in the superclass, aiming to promote reuse~\cite{Fowler:2018}. Therefore, this operation also corresponds to our criteria for composite refactoring and we say we performed a {\sffamily \small Composite Pull Up Field}.

\bigskip

\begin{figure}[!th]
	\centering
	\vspace{0.2cm}
    \includegraphics[width=0.8\textwidth]{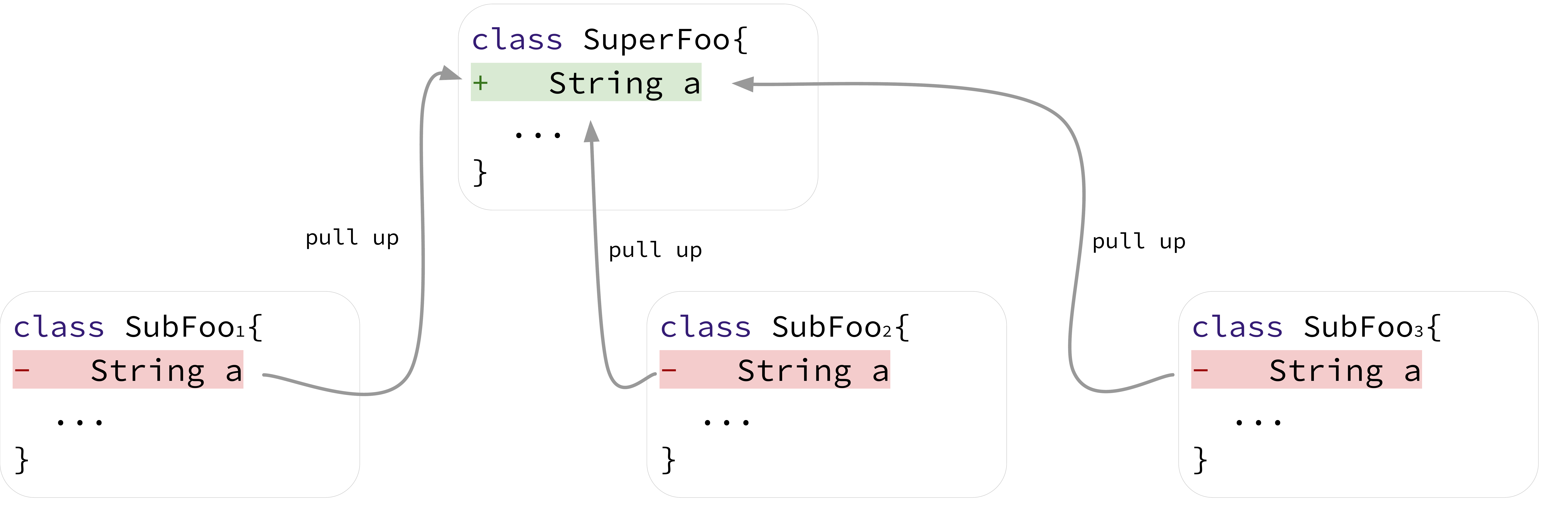}
	\caption{Composite Pull Up Field}
	\label{fig:fig_pull_up_decomposition_field}
	\vspace{0.2cm}
\end{figure}

\noindent {\em Mechanics:} First, we declare the field in the superclass. Then, we remove the declaration in the subclasses, as shown in Figure \ref{fig:fig_pull_up_decomposition_field}.
This operation can also be preceded by {\sffamily \small Rename Field}, aiming to standardize the names before the movement to the superclass.

\subsection{Composite Push Down Field}

\noindent {\em Motivation:} Similar to {\sffamily \small Push Down Method}, the goal involves moving data from a superclass to specific subclasses~\cite{Fowler:2018}. When this operation contemplates a sequence of fields movements, we say we performed a {\sffamily \small Composite Push Down Field}.

\bigskip

\begin{figure}[!th]
	\centering
    \includegraphics[width=0.55\textwidth]{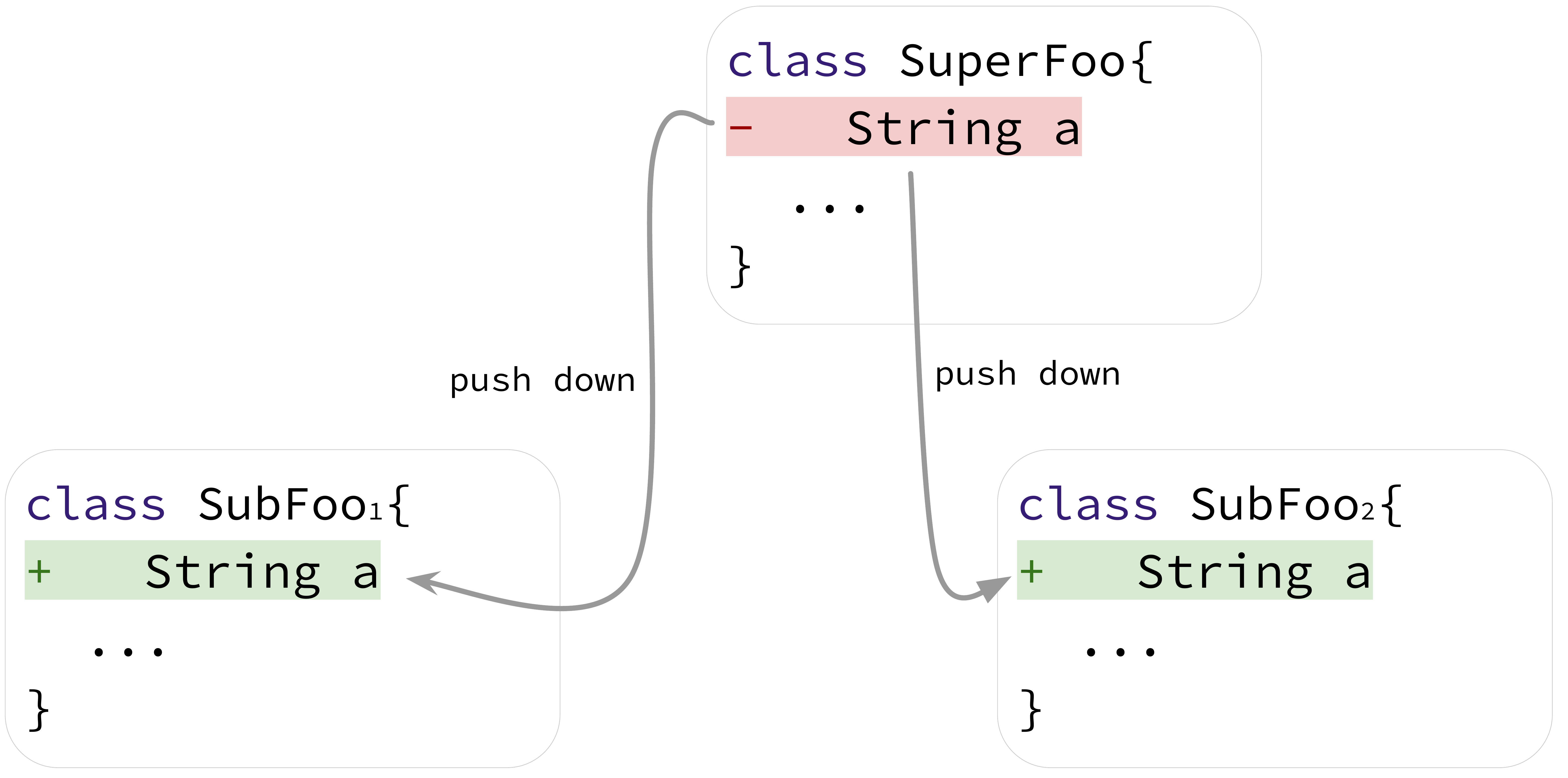}
	\caption{Composite Push Down Field}
	\label{fig:fig_push_down_decomposition_field}
\end{figure}

\noindent {\em Mechanics:} First, we declare the field in the required subclasses. Then, we remove the declaration in the superclass, as shown in Figure  \ref{fig:fig_push_down_decomposition_field}.

\subsection{A Final Note on Completeness}

The original catalog of refactorings proposed by Fowler has dozens of refactorings. Therefore, the  catalog of composites described in this section is much smaller (eight composites). On the one hand, this difference is expected because composites are coarse-grained and complex source code transformations, composed by atomic refactorings. On the other hand, it is also important to acknowledge that we do not claim on the completeness of the proposed catalog. Indeed, our central intention is to provide a comprehensive, well-documented, and easy to understand initial list of composite refactorings. In future studies, this list can be extended to include other types of composites.

\section{A First Oracle of Composite Refactoring}
\label{section:composite-oracle}

To investigate whether the proposed composite refactorings occur in real projects, we initially search for composite refactorings in one of the most representative refactoring oracles in the literature, curated by Tsantalis and other researchers~\cite{Tsantalis:2020:RefactoringMiner2,TsantalisOracle,Tsantalis:ICSE:2018:RefactoringMiner}. This oracle has been expanded over the years. The latest version includes more than 14K refactoring operations from 185 Java projects. The oracle instances were validated by multiple authors and/or well-known tools.  In other words, it is a trustworthy dataset for studying refactoring practices.

\subsection{Study Design}

\subsubsection{Research Questions Assessment}

We propose two research questions:

\bigskip

\noindent {\em (RQ1) What are the Most Common Composite Refactorings in the Oracle?} In the current version of the oracle, refactoring operations are reported as individual (i.e., non-composite) ones. Thus, in this first RQ, our goal is to explore the oracle data from a new perspective, looking for occurrences of composite refactorings. In other words, we aim to provide a new oracle view, which is not based on individual refactoring operations. For this purpose, we first compute the frequency (i.e., the number of occurrences) of each composite instance.

\bigskip

\noindent {\em (RQ2) What are the Characteristics of Composite Refactorings in the Oracle?} The rationale of this second research question is to understand the main characteristics of the composite refactorings detected in RQ1. Therefore, for each composite instance, we compute information such as its scope (i.e., location of the entities before and after a refactoring operation) and size (i.e., number of individual refactoring operations). 

\subsubsection{Dataset}

In January 2022, we retrieved the most recent oracle version. Then, {\bf we selected only  refactoring operations that could be part of composite operations}.\footnote{By construction, the discarded refactorings cannot be part of the composites included in our catalog. However, we acknowledge they can be part of future composites (in this case, therefore we will need to update the current oracle).} For example, the original oracle includes refactorings such as {\sffamily \small  Move Attribute} and {\sffamily \small  Rename Method}, which are not related at all with the composite refactorings described in Section \ref{section:catalog}.

\begin{figure}[!th]
	\centering
    \includegraphics[width=0.7\textwidth]{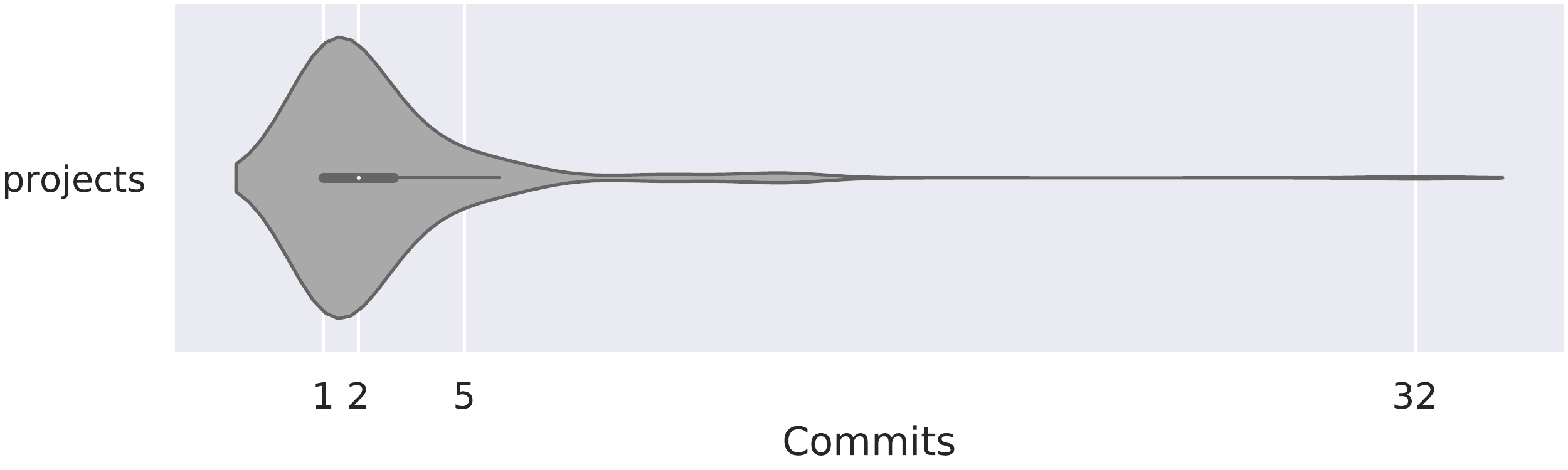}
	\caption{Distribution of commits per project (oracle)}
	\label{fig:fig_oracle_study_commits_per_project}
\end{figure}

As presented in Table \ref{table:operations-oracle}, our oracle sample includes 1,725 individual refactoring instances. 
Most instances are {\sffamily \small  Extract Method} (976 occurrences) and {\sffamily \small  Move Method} operations (227 occurrences). 
These operations are detected in 450 commits from 166 projects, such as \mcode{infinispan/infinispan} (a tool for storing, managing, and processing data)\footnote{\url{https://github.com/infinispan/infinispan}} and \mcode{gradle/gradle} (a build automation tool).\footnote{\url{https://github.com/gradle/gradle}}
Figure \ref{fig:fig_oracle_study_commits_per_project} shows the distribution of the number of selected commits per project.  As we can observe, the median is two commits, while the 90th percentile is about five commits. In the case of 78 projects (47\%), there are only refactoring instances from a single commit. In other words, the oracle sample does not include the whole project's history.

\begin{table}[!ht]
\vspace{-0.2cm}
\centering
\small
\caption{Selected refactoring operations in the oracle}
\label{table:operations-oracle}
\begin{tabular}{l r r r r}
\toprule
{\bf Operation} & {\bf Projects} & {\bf Commits} & {\bf Occurrences} & {\bf \%} 
\\ \midrule
Extract Method	& 140 & 329 & 976 & 56.6\\
Move Method	& 53 & 73 & 227 & 13.2\\
Inline Method & 48 & 64 & 127 & 7.4\\
Move and Rename Method	& 29 & 35 & 116 & 6.7\\
Extract and Move Method	& 29 & 35 & 114 & 6.6\\
Pull Up Method	& 24 & 28 & 74 & 4.3\\
Push Down Method & 10 & 11 & 30 & 1.7\\
Pull Up Field & 14 & 14 & 36 & 2.1\\
Push Down Field & 11 & 11 & 25 & 1.4\\
\midrule
All	& 166 & 450 & 1,725 & 100\\
\bottomrule
\end{tabular}        
\end{table}

\subsubsection{Detecting Composite Refactorings}
\label{sec:scripts-detect-composite}

We implement a set of scripts to detect the composite refactorings described in Section \ref{section:catalog}.  Their input comprises a list of individual refactoring operations. Basically, these scripts operate by searching for clusters of refactoring operations $R_1$, $R_2$,.., $R_n$ that can be replaced by a single composite refactoring {\em CR}. 
Therefore, we iterate over the list of refactorings detected in a system, grouping operations by considering the criteria described in Table \ref{table:detect-composite-condition}.

\begin{table}[!ht]
\centering
\footnotesize
\caption{Conditions to cluster two refactoring operations ($r_1$ and $r_2$) into a composite
}
\label{table:detect-composite-condition}
\begin{tabular}{l P{8.3cm}}
\toprule
 {\textbf{Composite}} &  {\textbf{Condition}} 
\\ 
\midrule
Method Composition & $signature(r1.target) = signature(r2.target) \wedge type(r1.target) = type(r2.target) \wedge (r1.refType, r2.refType) \in \{extract, extract\_move\}$\\\\[-0.25cm]
Method Decomposition & $signature(r1.source) = signature(r2.source) \wedge type(r1.source) = type(r2.source) \wedge (r1.refType, r2.refType)  \in \{extract, extract\_move\}$\\\\[-0.25cm]
Class Decomposition	& $type(r1.source) = type(r2.source) \wedge (r1.refType, r2.refType) \in \{move, move\_rename\}$\\\\[-0.25cm]
Composite Inline Method & $signature(r1.source) = signature(r2.source) \wedge type(r1.source) = type(r2.source) \wedge (r1.refType, r2.refType) \in \{inline\}$\\\\[-0.25cm]
Composite Pull Up Method & $signature(r1.target) = signature(r2.target)\wedge type(r1.target) = type(r2.target) \wedge 
 (r1.refType, r2.refType) \in \{pull\_up\}$\\\\[-0.25cm]
Composite Push Down Method & $signature(r1.source) = signature(r2.source) \wedge type(r1.source) =  type(r2.source) \wedge (r1.refType, r2.refType)  \in \{push\_down\}$\\\\[-0.25cm]
Composite Pull Up Field & $name(r1.target) = name(r2.target) \wedge \hspace{1.2cm} type(r1.target) = type(r2.target)  
\wedge (r1.refType, r2.refType) \in \{pull\_up\}$\\\\[-0.25cm]
Composite Push Down Field & $name(r1.source) = name(r2.source) \wedge type(r1.source) = type(r2.source) \wedge (r1.refType, r2.refType) \in \{push\_down\}$\\
\bottomrule
\end{tabular}        
\end{table}

For {\sffamily \small  Method Decomposition}, {\sffamily \small  Class Decomposition}, {\sffamily \small  Composite Push Down Method}, {\sffamily \small  Composite Push Down Field}, and {\sffamily \small  Composite Inline Method} (i.e., operations that break down code elements), we search for groups of refactorings that have as source the same code element.  For Composite {\sffamily \small  Pull Up Method}, {\sffamily \small  Composite Pull Up Field}, and {\sffamily \small  Method Composition}), we look for refactorings that have as target the same code element.

Moreover, the source and target checking vary according to each composite refactoring. For composites at the level of methods, we verify the signature and the class. For example, for {\sffamily \small Method Composition}, we group refactorings $r_1$ and $r_2$  into the same composite whenever the signature of the target methods are the same (i.e., $signature(r1.target) = signature(r2.target)$) and the target methods are in the same class (i.e., $type(r1.target) = type(r2.target)$).  We also check the respective refactoring types. In the case of {\sffamily \small Composite Pull Up Field} and {\sffamily \small  Composite Push Down Field}, i.e, composites at the level of fields, we verify the field's name and their respective class. Finally, for {\sffamily \small  Class Decomposition}, we group {\sffamily \small  Move Method} operations that originated from the same class  (i.e., $type(r1.source) = type(r2.source)$).

It is also important to mention that our criteria for grouping refactoring operations do not include time constraints. Therefore, two or more refactorings can be part of the same composite, even though they were performed in distinct periods over the system's history. We made this decision motivated by two considerations. First, it is not trivial to set a threshold for the duration of the composites. Second, because our main goal is to propose a catalog of composites, as well as to mine and analyze examples of these refactorings, even if they were performed in long time intervals.

After their execution, our scripts produce a list of composite refactorings, including a graph-based visualization and textual data. To validate the results, we manually inspected a sample of composite refactorings from the oracle. Specifically, we execute the following steps for each composite type:

\begin{enumerate}
    \item We selected a random sample of four instances (of each composite).
    \item For each selected instance:
    \begin{itemize}
        \item We carefully analyzed the respective refactorings in the oracle, verifying whether the operations are correct. In other words, we check the refactoring type, source, and target, as well as basic information such as project name and commit.
        \item In the last step, we verify if there are missing refactorings. In other words, we check if there are operations in the oracle that should be a part of the selected composite. For example, in \mcode{neo4j}, we detected a {\sffamily \small Method Composition} that creates the method \mcode{createCountsTracker()} in class \mcode{CountsComputerTest}.\footnote{\url{github.com/alinebrito/composite-refactoring-catalog/blob/main/results/oracle/neo4j/neo4j/results/composition_extract_method/view/subgraph_atomic_4.md}}  For this case, we verify if there are extractions to the same target that were not properly detected by our scripts.
    \end{itemize}
\end{enumerate}

We manually inspected 28 composites, since for {\sffamily \small Composite Push Down Method} and {\sffamily \small Composite Push Down Field}, we  detected only four instances in the oracle. Table \ref{table:scripts-selected-sample} summarizes the results. The size of the selected composites ranges from 2 to 39 refactoring operations, covering 160 refactorings from the oracle.   Overall, we did not identify errors by inspecting the sample of composite refactorings. In other words, we do not detect absent refactoring operations, i.e, operations that were not clustered correctly by our scripts. The scripts and inspected sample are publicly available at \url{github.com/alinebrito/composite-refactoring-catalog}.

\begin{table}[!ht]
\vspace{-0.2cm}
\centering
\small
\caption{Inspected sample of composite refactorings (Oracle)}
\label{table:scripts-selected-sample}
\begin{tabular}{l r r}
\toprule
{\bf Composite} & {\bf Instances} & {\bf Refactorings}
\\ \midrule
Method Composition & 4 & 48\\
Method Decomposition & 4 & 22\\
Class Decomposition & 4 & 52\\
Inline Method & 4 & 9\\
Pull Up Method & 4 & 11\\
Push Down Method & 2 & 4\\
Composite Pull Up Field & 4 & 10\\
Composite Push Down Field & 2 & 4\\
\midrule
All	& 28 & 160  \\
\bottomrule
\end{tabular}        
\end{table}

\subsection{Results}

\subsubsection{(RQ1) What are the Most Common Composite Refactorings in the Oracle?}

Among 1,725 single refactoring operations, an impressive number of 1,043 (60.5\%) are part of composite refactorings, as presented in Table \ref{table:oracle-rq1-frequency}. For example, 537 {\sffamily \small  Extract Method} or {\sffamily \small  Extract and Move Method} are part of {\sffamily \small  Method Composition} instances, which is the most frequent case. There are also significant rates of {\sffamily \small Method Decomposition} (125 occurrences, 34.1\%) and {\sffamily \small Class Decomposition} (55 occurrences, 15\%).
However, composite refactorings in the inheritance hierarchy are infrequent. For example, there are only 15 composite refactorings formed by {\sffamily \small Push Down Method} and {\sffamily \small Pull Up Method} operations (4.1\%). Also, there are a few occurrences of composites at the field level.

\bigskip

\begin{table}[!ht]
\centering
\small
\caption{Frequency of composite refactorings (Oracle)}
\label{table:oracle-rq1-frequency}
\begin{tabular}{l c c c r r}
\toprule
\multirow{2}{*}{\bf Name} & \multirow{2}{*}{\bf Projects} & \multicolumn{3}{c}{{\bf Composites}}
\\ 
\cline{3-5}
& & {\bf Operations} & {\bf Occurrences} & {\bf \%}\\
\midrule
Method Composition & 37 & 537 & 142 & 38.8\\
Method Decomposition & 37 & 295 & 125 & 34.1\\
Class Decomposition	& 37 & 277 & 55 & 15.0\\
Composite Inline Method & 11 & 48 & 21 & 5.7\\
Composite Pull Up Method & 7 & 33 & 13 & 3.6\\
Composite Push Down Method & 2 & 4 & 2 & 0.6\\
Composite Pull Up Field & 4 & 15 & 6 & 1.6\\
Composite Push Down Field & 1 & 4 & 2 & 0.6\\
\midrule
All & 81 & 1,043 & 366  & 100\\
\bottomrule
\end{tabular}        
\end{table}

Overall,  we detected 366 composite refactorings over 81 distinct projects. 
In 39 projects (48.1\%), there is only a single composite instance.
We identify most cases in a project called \mcode{Robovm}---127 instances grouping 389 single refactoring operations. 
Interestingly, this project also includes the largest composite refactoring instance, involving the composition of a method \mcode{has(...)}, which was created as the result of 30 {\sffamily \small  Extract Method} operations.\footnote{\url{https://github.com/robovm/robovm/commit/bf5ee44b}} The new method contains only a single line of code:

\smallskip

\begin{verbatim}
public boolean has(CFString key) {
    return data.containsKey(key);
}
\end{verbatim}

Therefore, this particular case of {\sffamily \small  Method Composition} was performed to remove code duplication (in this case, represented by a single line of code). It is worth mentioning that in the original oracle, this information was diluted over 30 individual and disconnected refactoring operations. By contrast, in our oracle view, they are represented by a single composite refactoring.

\begin{tcolorbox}[left=0mm,right=0mm,boxrule=0.25mm,colback=gray!5!white]
\vspace{-0.2cm}
{Summary of RQ1:} Out of 1,725 single refactoring operations, approximately 60\% are part of composite refactorings. We detected the instances of composite refactoring in 81 projects. The most recurring cases are Method Composition (142 occurrences, 38.8\%), Method Decomposition (125, 34.1\%), and Class Decomposition (55, 15\%).
\end{tcolorbox}

\subsubsection{(RQ2) What are the Characteristics of Composite Refactorings in the Oracle?}

\begin{figure}[!th]
	\centering
    \includegraphics[width=.6\textwidth]{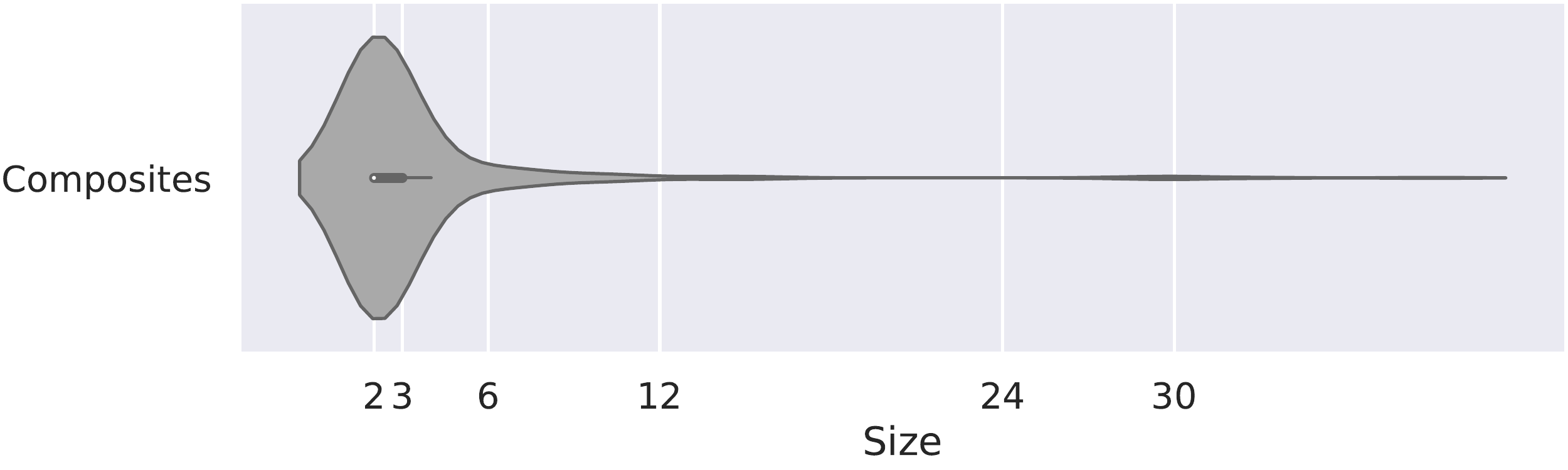}
	\caption{Distribution of the size of composite refactorings (Oracle)}
	\label{fig:fig_violin_composite_size_oracle}
\end{figure}

We also investigate the main characteristics of the composite refactorings detected in RQ1 in terms of size and scope. Regarding their size, i.e., the number of refactoring operations, most instances are small, as expected. As we can observe in Figure \ref{fig:fig_violin_composite_size_oracle}, about 84\% of the detected composite refactorings have up to three refactoring operations (308 occurrences). The values range from 2 to 39 refactoring operations per composite.

Next, we detail the results for the most important composites:

\smallskip

\noindent {\bf Class Decomposition.} Among the 55 instances of {\sffamily \small Class Decomposition}, 61.8\% refers to classes losing up to two methods (29 occurrences) or three methods (5 occurrences).  However, this category  includes one of the largest composite refactorings in the oracle, where a developer from \mcode{Graphhopper} decomposed a class by moving 39 methods.\footnote{\url{https://github.com/graphhopper/graphhopper/commit/7f80425b }} Interestingly, in the commit message, the developer added a brief description regarding the motivation, which is related to a well-known design principle (use composition instead of inheritance~\cite{GoF:Book:1994}): 

\medskip

\noindent {{``\em Refactoring of [class name]: use composition instead of inheritance''}}\\[-0.2cm]

\begin{figure}[!th]
	\centering
    \includegraphics[width=.56\textwidth]{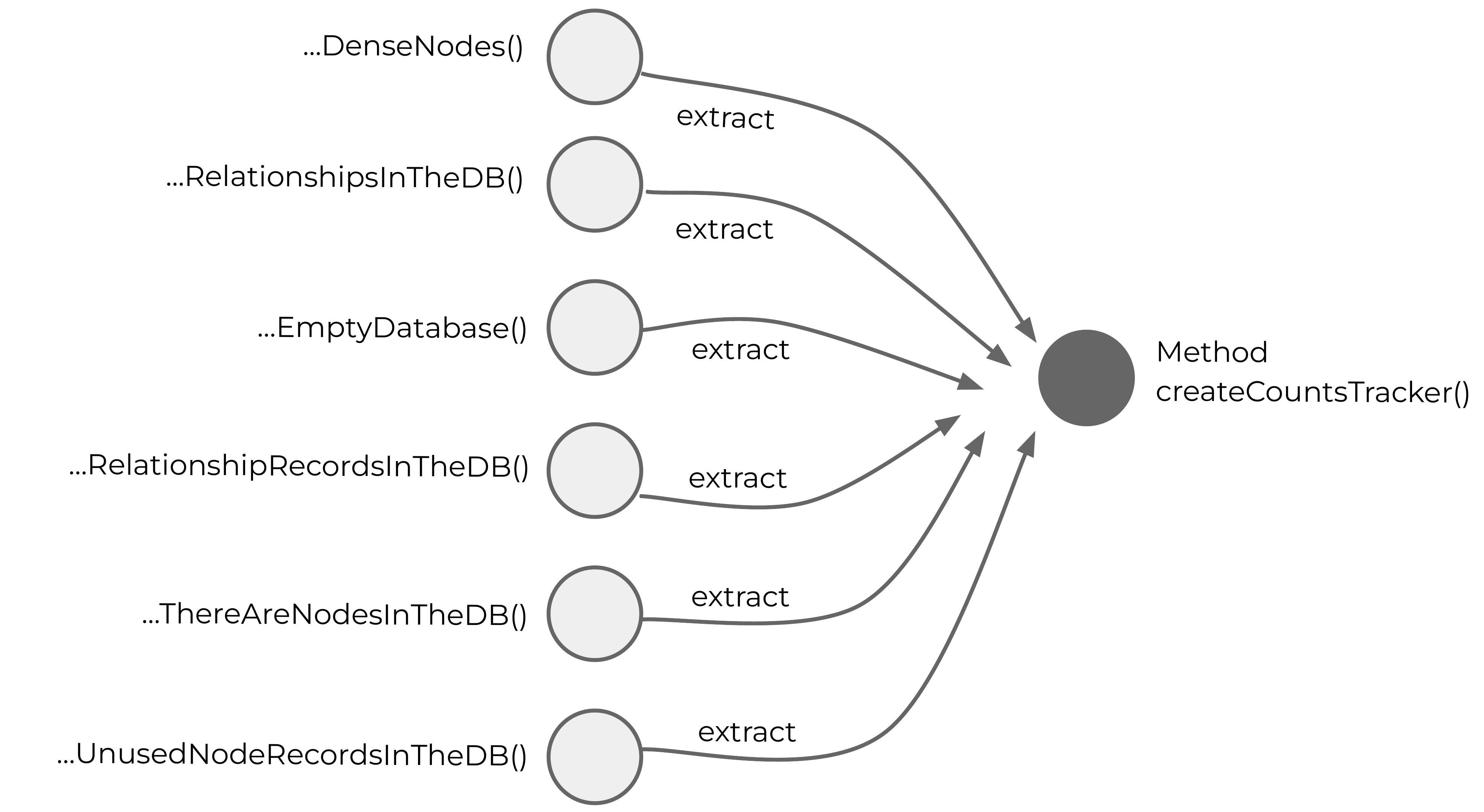}
	\caption{Example of Method Composition from Neo4j  (Oracle)}
	\label{fig:neo4j_composition_method_4}
\end{figure}

\noindent {\bf Method Composition.} The size of the 142 instances of {\sffamily \small  Method Composition} varies from 2 to 31 operations, with a median of two operations per composite. Furthermore, most {\sffamily \small Method Composition} instances are {\em intra-class} (121 occurrences, 85\%), i.e., the source methods are located in the same class of the target method. Figure \ref{fig:neo4j_composition_method_4} shows an example from \mcode{Neo4j}, where a developer extracted a method called \mcode{createCountsTracker()} from six methods.\footnote{\url{https://github.com/neo4j/neo4j/commit/5fa74fbb}} All refactorings happened in the scope of the same class \mcode{CountsComputerTest}. However, in the original oracle, these refactorings are reported as six distinct and unrelated operations. Finally, in 10 cases (7\%), the composites are {\em inter-class}, i.e.,  developers compose methods by ``merging'' pieces of code coming from distinct classes. The remaining are {\em mixed} {\sffamily \small  Method Decomposition}, including the two categories.

\bigskip

\medskip

\noindent {\bf Method Decomposition.} 95\% of the instances of {\sffamily \small  Method Composition} (119 occurrences), which were detected in 37 projects, have up to three operations. The values range from 2 to 15 operations, distributed among {\em intra-class} cases (91\%), {\em inter-class} cases (3\%), and {\em mixed} ones (6\%). 

\bigskip

\noindent {\bf Other cases.} In the oracle, there are only 127 occurrences of {\sffamily \small Inline Method}. Consequently, we also found a few cases of composite inlines (21 instances, 5.7\%), including at most four operations. The same applies to composite refactorings over inheritance hierarchies: {\sffamily \small  Composite Pull Up Method} (13 instances, 3.6\%), {\sffamily \small  Composite Push Down Method} (2 instances, 0.6\%), {\sffamily \small  Composite Pull Up Field} (6 instances, 1.6\%), and {\sffamily \small  Composite Push Down Field} (2 instances, 0.6\%).

\begin{tcolorbox}[left=0mm,right=0mm,boxrule=0.25mm,colback=gray!5!white]
\vspace{-0.2cm}
{Summary of RQ2:} Most composite refactorings are small, including up to three operations. However, we also detect large instances, for example, 30 {\sffamily \small  Extract Method} operations to compose a single method. Regarding the scope of the operations, most {\sffamily \small   Method Composition} and {\sffamily \small  Method Decomposition} are {\em intra-class}. In other words, developers usually extract multiple methods to the current classes.
\end{tcolorbox}

\section{Composite Refactoring in the Wild}
\label{section:composite-wild}

In the first study, we look for composites in a well-known oracle. However, the refactoring instances selected for this oracle do not cover the complete history of each project. In other words, the oracle used in Section \ref{section:composite-oracle} only contains selected refactoring instances. Therefore, we might have missed operations in the reported composite refactorings simply because they were not selected for inclusion in the oracle. To tackle this issue, we decided to perform a complementary study, in which we search for composite refactorings in the complete history of ten popular GitHub-based projects.\footnote{Since operations at the field level are infrequent, and it is also unsupported by the current RefDiff tool version, we decide not to include them in this complementary study.}

\subsection{Study Design}

\subsubsection{Research Questions Assessment}

As in the  study described in Section \ref{section:composite-oracle}, we propose two research questions:

\bigskip

\noindent {\em (RQ3) What are the Most Common Composite Refactorings in the Wild?} Similarly to RQ1, we assess the frequency of each composite refactoring, but now in 10 popular GitHub projects. \bigskip

\noindent {\em (RQ4) What are the Characteristics of Composite Refactorings in the Wild?} Similarly to RQ2, the rationale of this research question is to shed light on the main characteristics of composite refactorings while considering the complete development history of 10 projects.

\subsubsection{Dataset}

To answer the proposed research questions, we relied on a set of real-world and popular projects. Specifically, we selected the top-10 Java projects on GitHub, ordered by their number of stars. We adopted this criterion because stars is a relevant metric to identify popular repositories~\cite{jss-2018-github-stars,hudson:icsme2016:Popularity}. Moreover, in our sample, we only include projects that are software systems. For example, despite having a high number of stars, we did not include \textit{kdn251/interviews} (a guide for interviews),\footnote{\url{https://github.com/kdn251/interviews}} and \textit{iluwatar/java-design-patterns} (a set of code samples).\footnote{\url{https://github.com/iluwatar/java-design-patterns}} Table \ref{table:selected-projects} describes the selected projects, including basic information, such as number of stars, commits, contributors, and short descriptions. The selected projects are from distinct domain areas, including web frameworks and animation libraries.

\begin{table}[!th]
\centering
\small
\caption{Selected Java projects}
\label{table:selected-projects}
\begin{tabular}{l r r r l}
\toprule
{\bf Project} & {\bf Stars} & {\bf Comm.} & {\bf Contr.} & {\bf Short Description}
\\ \midrule

Spring Boot & 56,717 & 33,692 &  831 & Support framework  \\ 
Elasticsearch & 56,081& 60,227 & 1,651 & Analytics engine  \\ 
RxJava & 45,055 &  5,921 & 278 &Event-based library  \\ 
Spring Framework & 43,943 & 22,728 &  551 & Support framework   \\
Google Guava & 42,045 & 5,609 & 265 & Core Java libraries \\
Square Retrofit & 38,539 & 1,902 & 158& HTTP client  \\ 
Apache Dubbo & 35,968 & 4,848 &  349 &RPC framework  \\ 
MPAndroidChart & 33,811 & 2,070 &  69 & Chart library  \\ 
Lottie Android & 31,612 & 1,321 &  106 & Rendering library  \\ 
Glide & 31,578 & 2,592 & 131 & Image library \\
\bottomrule
\end{tabular}  
\vspace{-0.25cm}
\end{table}

\subsubsection{Detecting Composite Refactorings}

To detect composite refactorings, we need first to identify  single refactoring operations. For this purpose, we used RefDiff, a well-known multi-language refactoring tool~\cite{danilo:msr2017:RefDiff, danilo:tse2020:refdiff2}. 
As usual in git-based mining tools, RefDiff detects refactorings by comparing a commit with its parent commit.  To facilitate the usage of the tool, we first implemented a set of scripts that automate tasks such as downloading GitHub projects and retrieving the list of commits from the default branch. 
The scripts then rely on RefDiff to detect single refactoring operations. They also automatically exclude refactorings in non-core packages, such as \textit{``test''} and \textit{``sample''}. 
The final step concerns the detection of the composites defined in our catalog, using the scripts  described in Section \ref{section:composite-oracle}.

\subsection{Results}

\subsubsection{(RQ3) What are the Most Common Composite Refactorings in the Wild?}

As presented in Table \ref{table:rq3-frequency}, we identify 2,886 occurrences of composite refactorings.
Most cases refer to {\sffamily \small Class Decomposition} (957 occurrences, 33.2\%), i.e., 957 classes and interfaces have lost multiple methods. The values range from 8 classes in {\sffamily \small Lottie} {\sffamily \small Android} to 280 classes in {\sffamily \small Elasticsearch}.

\begin{table}[!ht]
\vspace{-0.2cm}
\centering
\small
\caption{Frequency of composite refactorings (in the wild)}
\label{table:rq3-frequency}
\begin{tabular}{l r r r r r}
\toprule
\multirow{2}{*}{\textbf{Name}} &  \multicolumn{2}{c}{\textbf{Wild}} & & \multicolumn{2}{c}{\textbf{Oracle}}\\
\cline{2-3} \cline{5-6}
& {\bf Occur.} & {\bf \%} & & {\bf Occur.} & {\bf \%} \\
\midrule

Class Decomposition	& 957 & 33.2 & & 55 &  15.0\\

Method Decomposition & 683	& 23.7 & & 125 & 34.1 \\

Method Composition	&582	& 20.2 & & 142 & 38.8 \\

Composite Pull Up Method	& 450	& 15.6 & & 13 & 3.6 \\

Composite Inline Method & 129	& 4.5 & & 21 & 5.7 \\

Composite Push Down Method & 85 & 2.8 & & 2 &  0.6 \\

Composite Pull Up Field &  - & - & & 6 & 1.6 \\

Composite Push Down Field & -  & - & & 2 &  0.6 \\

\midrule

All	& 2,886 & 100 & & 366 & 100 \\

\bottomrule
\end{tabular}        
\end{table}

Moreover, about 32\% of the composites are from {\sffamily \small Elasticsearch}, a popular search engine.\footnote{\url{https://github.com/elastic/elasticsearch}} In this project, we detect  921 composites grouping 3,310 single refactoring operations.  Among them, most cases refer to {\sffamily \small Class Decomposition} (280 occurrences, 30.4\%).

\begin{figure}[!ht]
	\centering
    \includegraphics[width=0.45\textwidth]{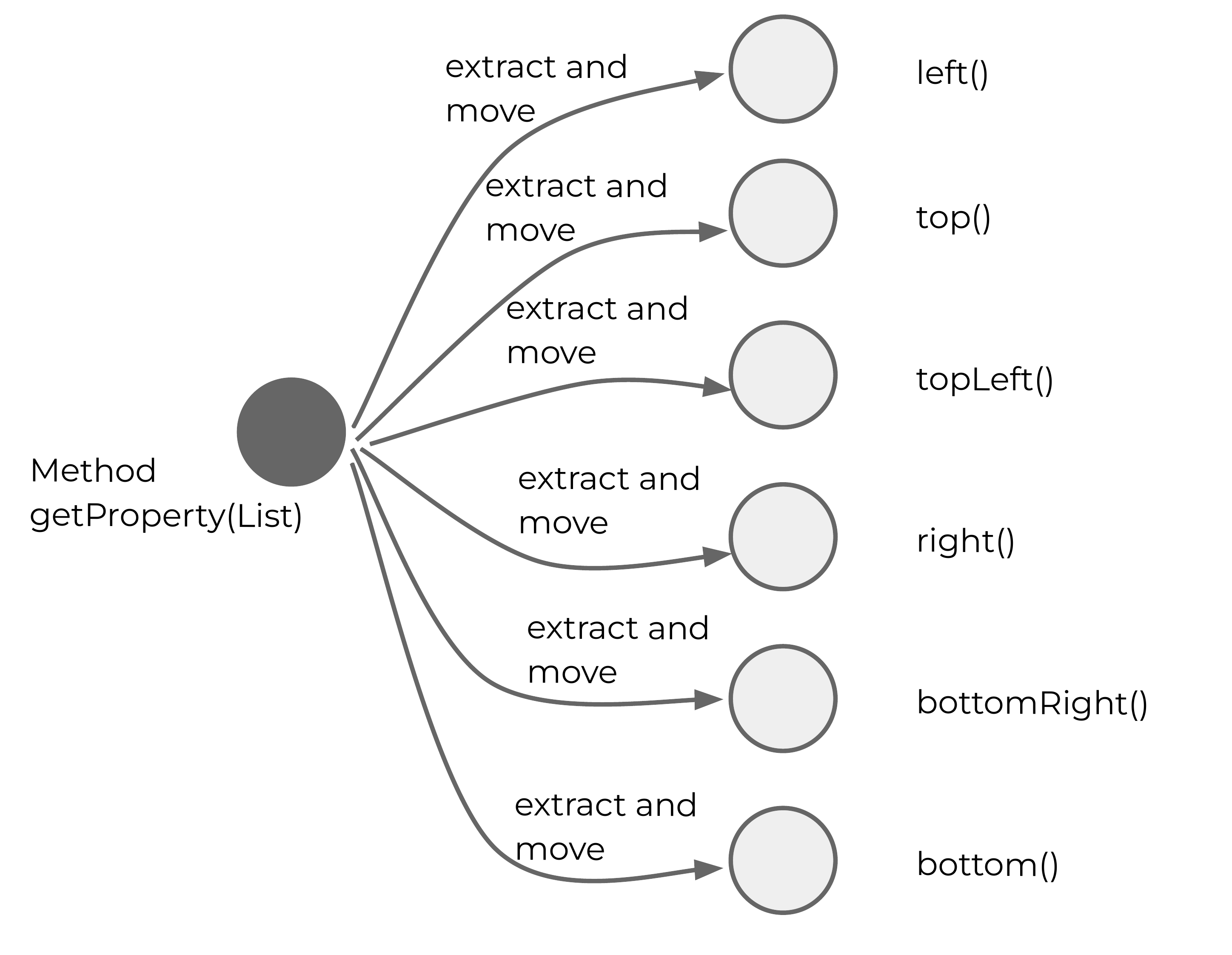}
	\caption{Example of Method Decomposition in Elasticsearch}
	\label{fig:dec_extract_subgraph_atomic_139}
\end{figure}

There is also a significant number of {\sffamily \small Method Decomposition} (683 occurrences, 23.7\%), such as in the example of Figure \ref{fig:dec_extract_subgraph_atomic_139}. 
In this case,  method \mcode{getProperty(List)} lost multiple pieces of code, after a developer performed six {\sffamily \small Extract and Move} operations in a single commit.

Interestingly, all extracted methods were moved to the same class \mcode{GeoBoundingBox}. In the commit description,\footnote{\url{https://github.com/elastic/elasticsearch/commit/769650e0}} the maintainer points out the intention to centralize related  logic: \\[-0.2cm]

\noindent{{\em
``A lot of this logic can be centralized
instead of having separated efforts to do the same things''}

\bigskip

\begin{tcolorbox}[left=0mm,right=0mm,boxrule=0.25mm,colback=gray!5!white]
\vspace{-0.2cm}
{\em Summary of RQ3:} In our extended dataset, the most common composite refactorings are {\sffamily \small Class Decomposition} (957 occurrences, 33.2\%); 
{\sffamily \small Method Decomposition} (683  occurrences, 23.7\%); 
and {\sffamily \small Method Composition} (582 occurrences, 20.2\%).
There are also a few occurrences of composite refactorings related to inheritance, i.e., {\sffamily \small Composite Pull Up Method} and {\sffamily \small Composite Push Down Method}.
\end{tcolorbox}

\noindent{\em Comparison with the oracle results (RQ1):} In Table \ref{table:rq3-frequency}, we also report the results obtained with the oracle sample, aiming to facilitate  comparison. As we can notice, the frequency of composites follows a similar tendency, i.e., the top-3 cases are exactly the same: {\sffamily \small Class Decomposition}, {\sffamily \small Method Decomposition}, and {\sffamily \small Method Composition}. However, in the oracle, the order is the reverse  (e.g., {\sffamily \small Method Composition} is the most frequent composite).

\subsubsection{(RQ4) What are the Characteristics of Composite Refactorings in the Wild?}

Regarding their size---as measured by the number of single refactorings in each composite---most instances in the extended dataset are also small. 
Figure \ref{fig:fig_violin_composite_size} presents the size distribution per project, after removing outliers, since they tend to distort the plot's aspect. In all projects, the median is two or three operations. However, there are also large composites, for example, the largest case includes dozens of operations, in which several methods were moved from a single class.\footnote{\url{https://github.com/ReactiveX/RxJava/commit/10325b90}}  In the following paragraphs, we detail the characteristics and give examples of each composite refactoring.

\medskip

\begin{figure}[!th]
	\centering
    \includegraphics[width=1\textwidth,trim={4cm 0 3cm 0},,clip]{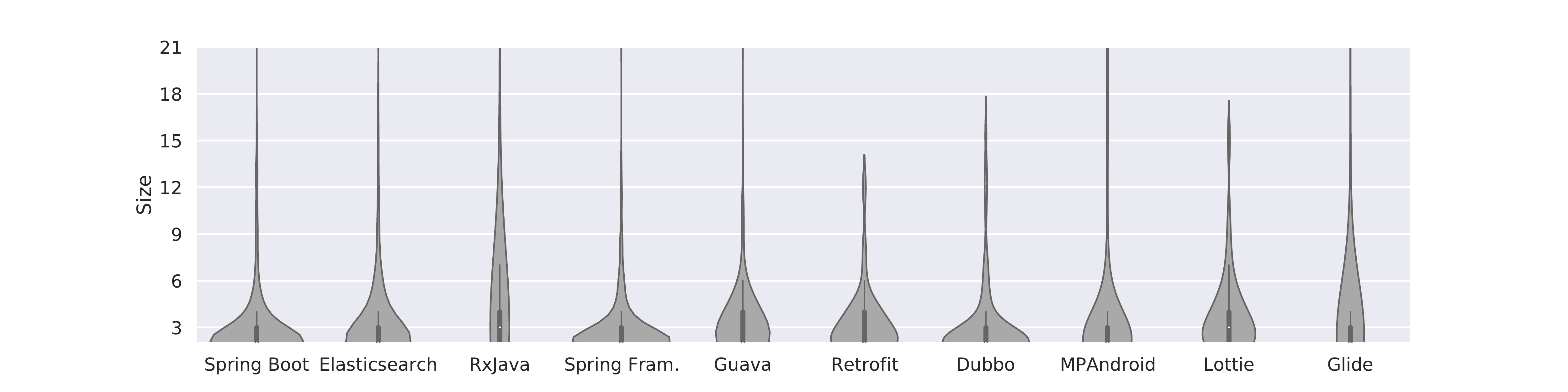}
	\caption{Distribution of the size of composite refactorings per project}
	\label{fig:fig_violin_composite_size}
\end{figure}

\noindent {\bf Class Decomposition.} 
 Figure \ref{fig:fig_degree_dec_move_method} summarizes the size results of  {\sffamily \small Class Decomposition}. As we can notice, most composites of this type are small.
 About 62\% of the cases involve up to three operations, such as in the example in Figure \ref{fig:dec_move_lottie_5}.
 In this example, a class of {\sffamily \small Lottie Android} lost three methods in two commits. However, there are also large instances. For example, in {\sffamily \small Google Guava} one developer moved each method from class \mcode{EmptyImmutableMap} to a distinct one, i.e., he performed a composite refactoring composed of ten operations.\footnote{\url{https://github.com/google/guava/commit/d8f98873}}

\begin{figure}[!th]
	\centering
    \includegraphics[width=0.55\textwidth,trim={0.5cm 1.5cm 1.5cm 1cm},,clip]{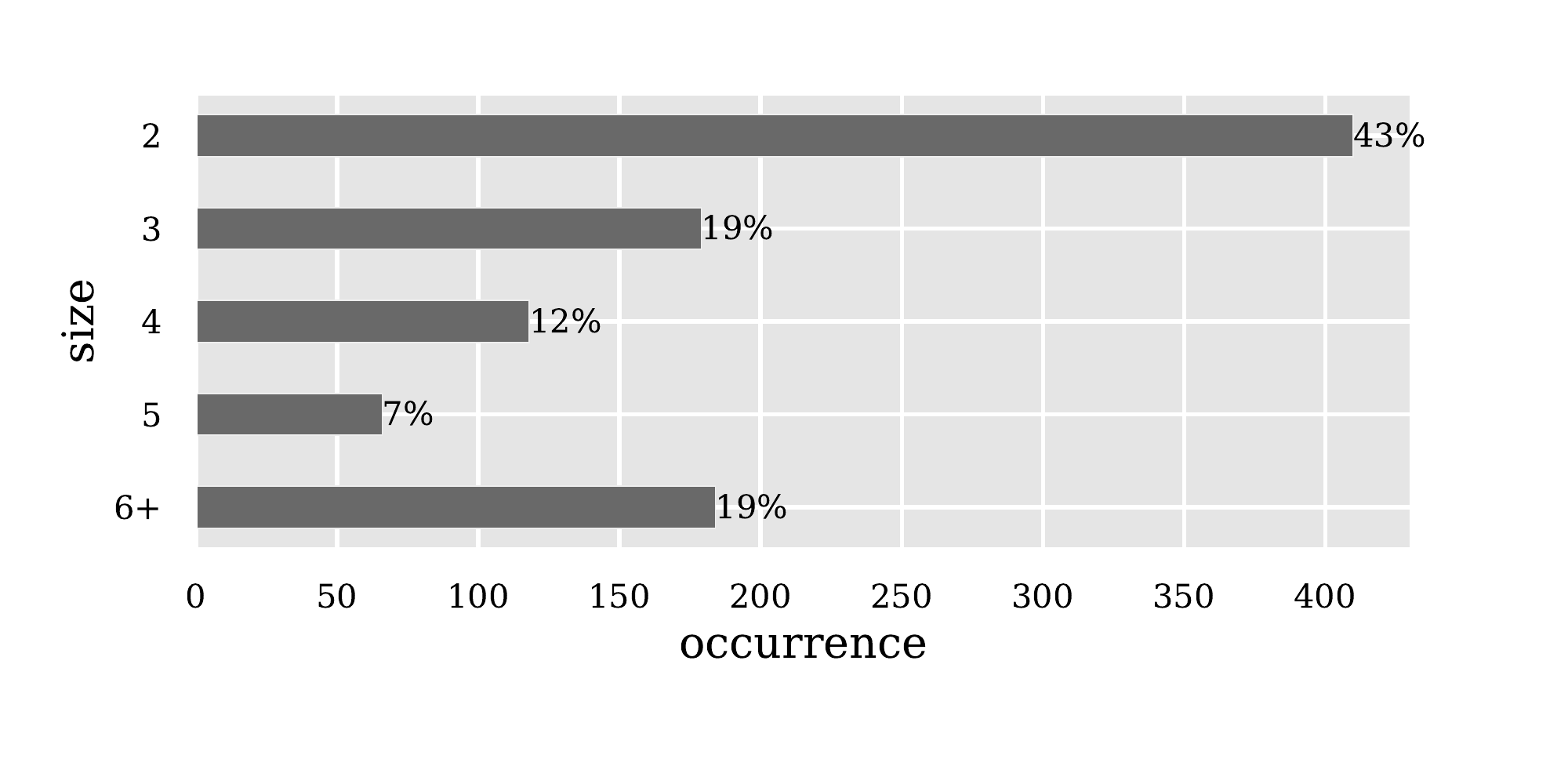}
	\caption{Number of operations by composite refactoring (Class Decomposition)}
	\label{fig:fig_degree_dec_move_method}
\end{figure}

\begin{figure}[!th]
	\centering
    \includegraphics[width=0.6\textwidth]{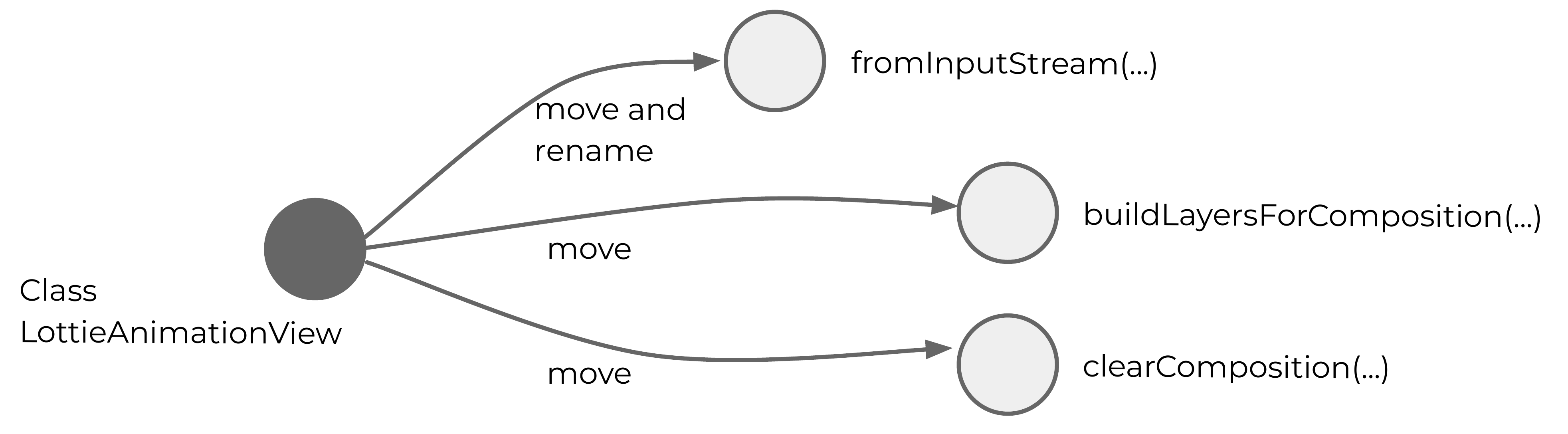}
	\caption{Example of Class Decomposition in Lottie Android}
	\label{fig:dec_move_lottie_5}
\end{figure}

\bigskip

\noindent {\bf Method Decomposition.} 
As in the study described in Section 4, we also separate the composites into {\em intra-class} (i.e., extractions to the same class of the fragmented method), {\em inter-class} (i.e., when the extracted methods are moved to distinct classes), and {\em mixed} (i.e., both cases), as shown in Table \ref{table:rq1-method-decomposition-location}.
As we can observe, most extractions are in the \textit{intra-class} category (317 occurrences, 46\%). However, another significant part of the results are \textit{inter-class} (238 occurrences, 35\%), i.e., all extracted methods are kept in the current class. 
We also investigate the number of extractions, i.e., the size of the {\sffamily \small Method Decomposition} instances.
As presented in Figure \ref{fig:fig_degree_dec_extract},  most  cases refer to methods decomposed using two {\sffamily \small Extract} operations (527 occurrences, 77\%) or three operations (108 occurrences, 16\%).
\\[-0.2cm]

\begin{table}[!ht]
\vspace{-0.25cm}
\centering
\small
\caption{Characteristics of Method Decomposition (in the wild)}
\label{table:rq1-method-decomposition-location}
\begin{tabular}{p{4cm} r r r r r r r}
\toprule
{\bf Project} & {\bf Occur.} &  \multicolumn{2}{c}{{\bf Intra-class}} & \multicolumn{2}{c}{{\bf Inter-class}} & \multicolumn{2}{c}{{\bf Mixed}}\\
\cline{3-8}
&  &  Occur. & \% &Occur. & \%  & Occur. & \% \\
\midrule

Spring Boot	& 148	&	107	&	72	&	21	&	14	&	20	&	14	\\

Elasticsearch	& 	234	&	75	&	32	&	118	&	50	&	41	&	18	\\

RxJava	& 	5	&	2	&	40	&	3	&	60	&	0	&	0	\\

Spring Framework	& 	152	&	77	&	51	&	38	&	25	&	37	&	24	\\

Guava	& 	10	&	3	&	30	&	5	&	50	&	2	&	20	\\

Retrofit	& 	6	&	2	&	33	&	2	&	33&	2	&	33	\\

Dubbo	& 	51	&	22	&	43	&	20	&	39	&	9	&	18	\\

MPAndroidChart	& 	35	&	5	&	14	&	25	&	71	&	5	&	14	\\

Lottie Android	& 	14	&	5	&	36	&	5	&	36	&	4	&	29	\\

Guide	& 	28	&	19	&	68	&	1	&	4	&	8	&	29	\\

\midrule

All	&  683	&	317	&	46	&	238	&	35	&	128	&	19	\\

\bottomrule
\end{tabular}        
\end{table}

\begin{figure}[!th]
	\centering
	\vspace{0.3cm}
    \includegraphics[width=0.65\textwidth,trim={0.5cm 0 1.5cm 0},,clip]{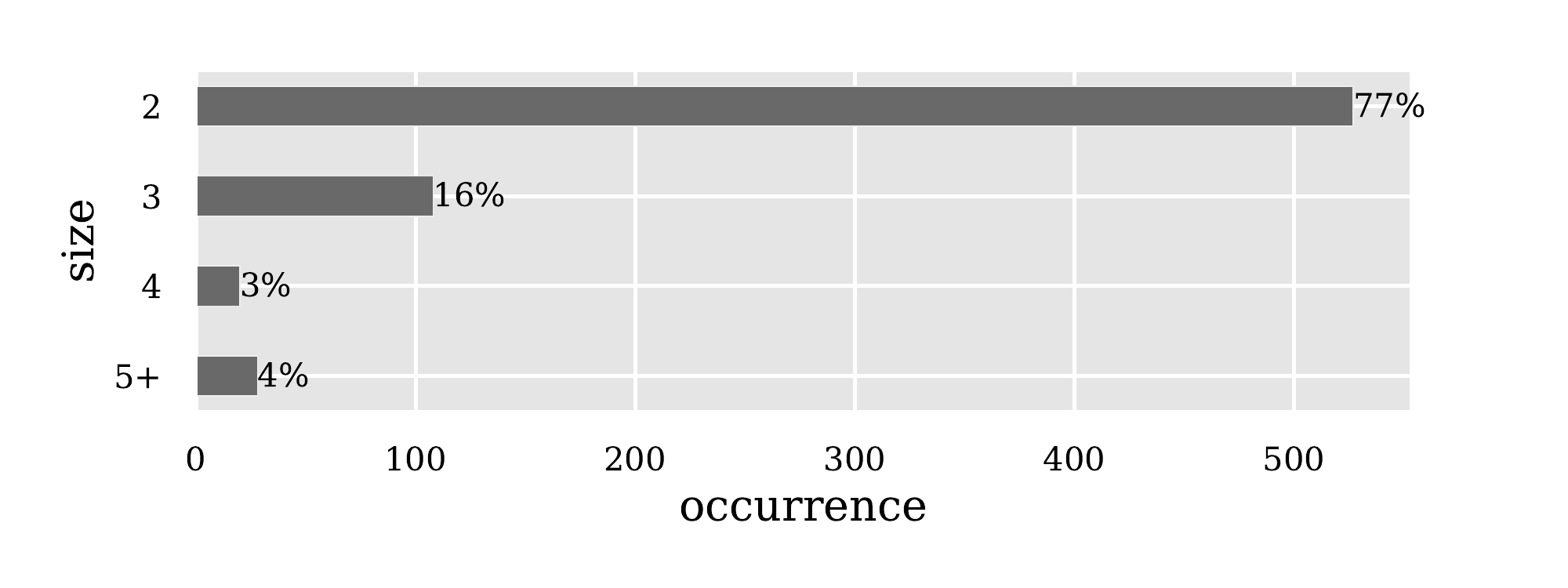}
	\caption{Number of operations by composite refactoring (Method Decomposition)}
	\label{fig:fig_degree_dec_extract}
\end{figure}

\noindent {\bf Method Composition.} Among  582 instances of this composite refactoring, frequently, the extracted code is moved to distinct classes (345 occurrences, 59\%), i.e., they are \textit{inter-class}, as shown in Table \ref{table:rq1-method-composition-location}.
Figure \ref{fig:fig_degree_com_extract_method} shows the results considering the size: as we can observe, most cases involve up to three operations (467 occurrences, 80\%).  {\sffamily \small Dubbo} includes an outlier, in which a developer extracted a utility method called \mcode{isEmptyMap(Map)} from seven other methods.\footnote{\url{https://github.com/apache/dubbo/commit/458a4504}} The extracted method has the following code:\\[-0.4cm]

\begin{small}
\begin{verbatim}
public static boolean isEmptyMap(Map map) {
    return map == null || map.size() == 0;
}
\end{verbatim}
\end{small}

\bigskip

\begin{table}[!ht]
\vspace{-0.3cm}
\centering
\caption{Characteristics of Method Composition (in the wild)}
\small
\label{table:rq1-method-composition-location}
\begin{tabular}{p{3.5cm} r r r r r r r}
\toprule
{\bf Project} & {\bf Occur.} &  \multicolumn{2}{c}{{\bf Intra-class}} & \multicolumn{2}{c}{{\bf Inter-class}} & \multicolumn{2}{c}{{\bf Mixed}}\\
\cline{3-8}
&  &  Occur. & \% &Occur. & \%  & Occur. & \% \\
\midrule
Spring Boot	&	79	&	34	&	43	&	42	&	53	&	3	&	4	\\
Elasticsearch	&	219	&	49	&	22	&	161	&	74	&	9	&	4	\\
RxJava	&	8	&	2	&	25	&	6	&	75	&	0	&	0	\\
Spring Framework &	151	&	67	&	44	&	66	&	44	&	18	&	12	\\
Guava	&	17	&	5	&	29	&	10	&	59	&	2	&	12	\\
Retrofit	&	5	&	1	&	20	&	4	&	80	&	0	&	0	\\
Dubbo	&	39	&	20	&	51	&	15	&	38	&	4	&	10	\\
MPAndroidChart	&	30	&	5	&	17	&	22	&	73	&	3	&	10	\\
Lottie Android	&	12	&	1	&	8	&	10	&	83	&	1	&	8	\\
Guide	&	22	&	13	&	59	&	9	&	41	&	0	&	0	\\

\midrule

All	&	582	&	197	&	34	&	345	&	59	&	40	&	7	\\

\bottomrule
\end{tabular}     
\vspace{-0.25cm}
\end{table}

\bigskip
\bigskip

\begin{figure}[!th]
	\centering
    \includegraphics[width=0.6\textwidth,trim={0.5cm 0 1.5cm 0},,clip]{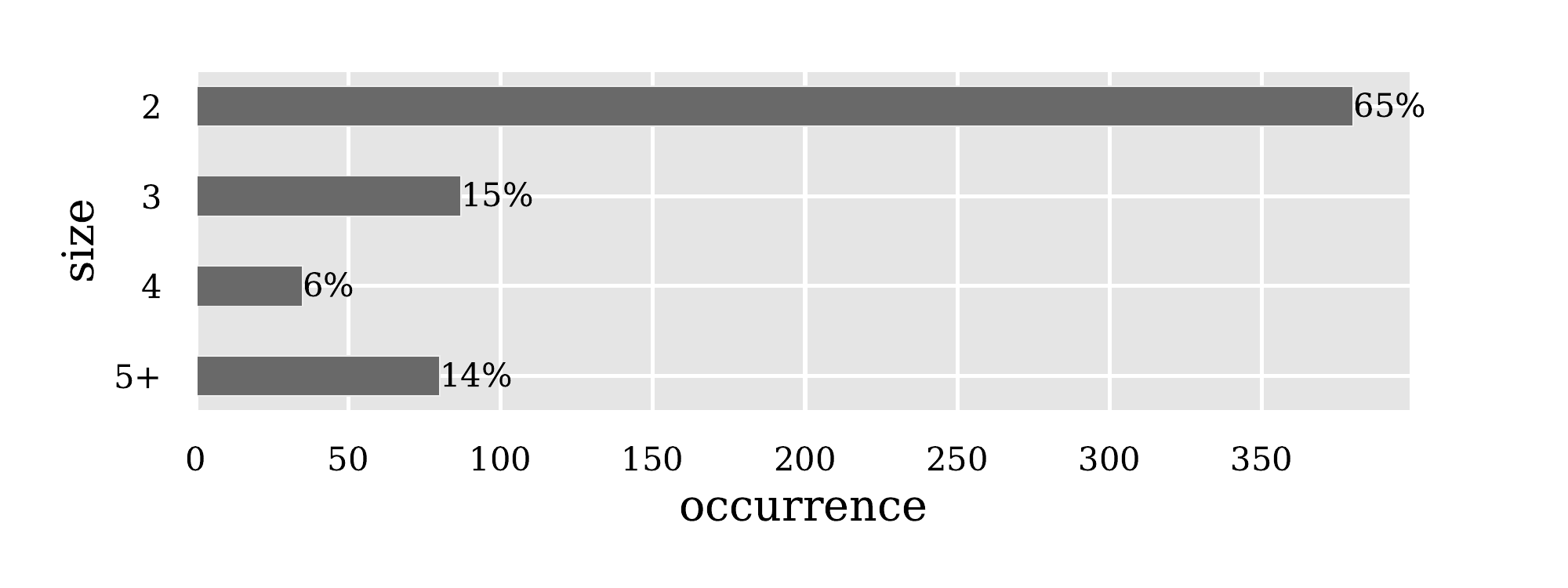}
	\caption{Number of operations by composite refactoring (Method Composition)}
	\label{fig:fig_degree_com_extract_method}
\end{figure}

\noindent {\bf Composite Pull Up Method.} In this category, the number of operations follows the same tendency detected in RQ2, e.g.,~most cases comprise two  (311 occurrences, 69\%) or three operations (65 occurrences, 15\%), as reported in Figure \ref{fig:fig_degree_com_pull_up}. However, 8\% of the occurrences have five or more operations.
{\sffamily \small Spring Boot} includes an example, in which a developer moved method \mcode{matches(...)} from five subclasses to the superclass \mcode{SpringBootCondition}.\footnote{\url{https://github.com/spring-projects/spring-boot/commit/840fdeb5}} In the commit description, the developer mentioned his intention, which relates to the improvement of the inheritance hierarchy.:
\\[-0.2cm]

\noindent{{\em
``Create common [name class] base class... This removes the need for [class name] and simplifies many of the existing condition implementations.''
}}
\\[-0.2cm]

\begin{figure}[!ht]
	\centering
	\vspace{-0.25cm}
    \includegraphics[width=0.55\textwidth,trim={0.5cm 0 1.5cm 0},,clip]{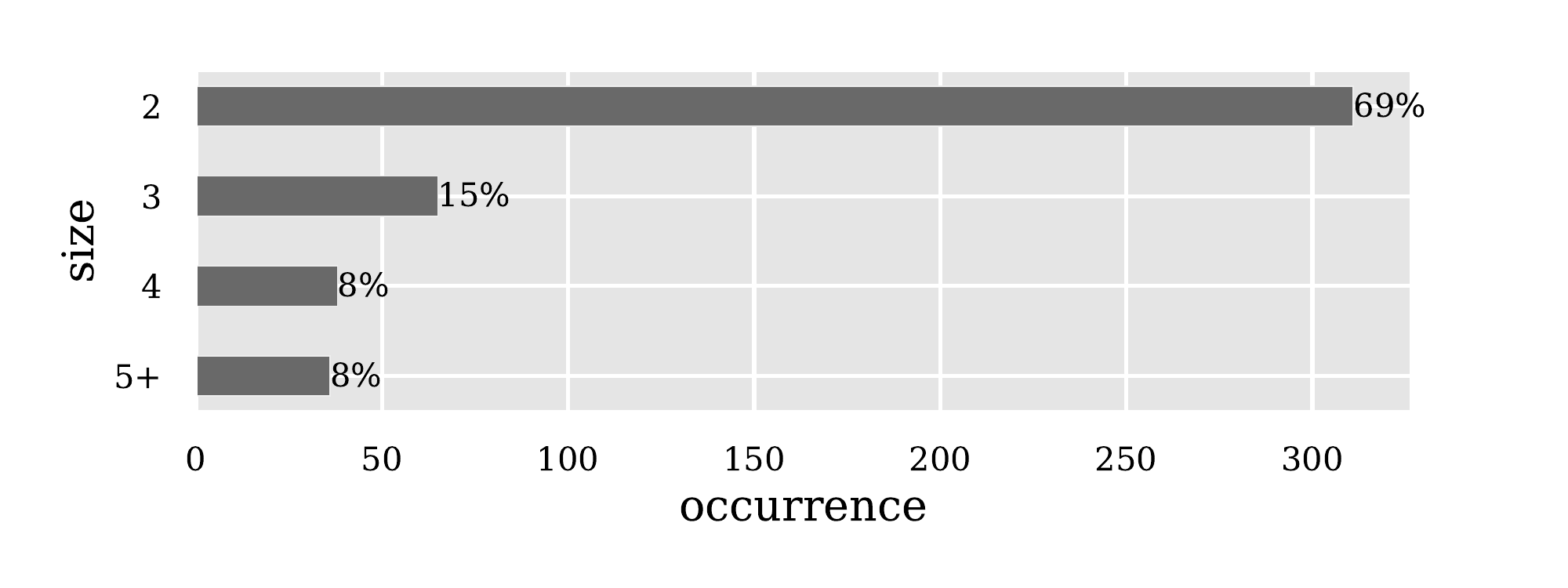}
	\caption{Number of operations by composite refactoring (Composite Pull Up Method)}
	\label{fig:fig_degree_com_pull_up}
\end{figure}

\noindent {\bf Composite Push Down Method.} 
Figure \ref{fig:fig_degree_dec_push_down} presents the results regarding the size of this type of composite refactoring. 
Overall, most cases comprise operations to move a  method to at most three subclasses (82 occurrences, 97\%). 

\begin{figure}[!th]
	\centering
    \includegraphics[width=0.55\textwidth,trim={0.5cm 0 1.5cm 0},,clip]{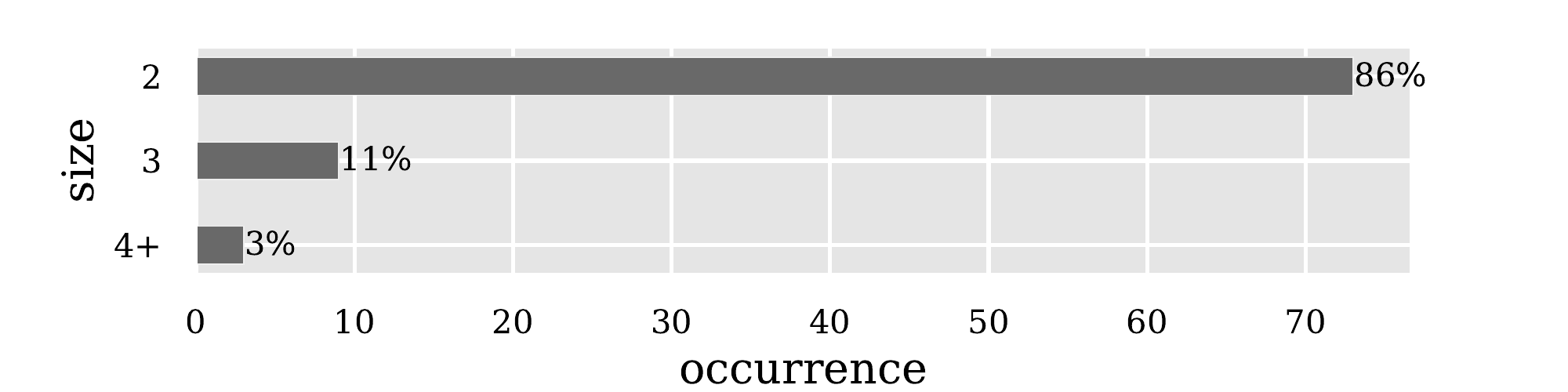}
	\caption{Size of composite refactorings (Composite Push Down Method)}
	\label{fig:fig_degree_dec_push_down}
	\vspace{-0.2cm}
\end{figure}

\bigskip

\noindent {\bf Composite Inline Method.} 
Regarding the number of affected elements,  most {\sffamily \small Composite Inline} operations involve two or three operations (109 occurrences, 85\%), as presented in Figure \ref{fig:fig_degree_dec_inline}.
{\sffamily \small Elasticsearch} includes a large instance, in which  a developer removed method \mcode{cast(Input, Output)} by performing 23 {\sffamily \small Inline Method} operations.\footnote{\url{https://github.com/elastic/elasticsearch/commit/022d3d7d}} 
In the commit description, the developer explained his motivation in the following way:\\[-0.2cm]

\noindent {{``\em Remove [functionality name] from [class name] as mutable state...  this is no longer necessary as each cast is only used directly in the semantic pass after its creation...''}}\\[-0.2cm]

\begin{figure}[!th]
	\centering
    \includegraphics[width=0.55\textwidth,trim={0.5cm 0.5 1.5cm 0},,clip]{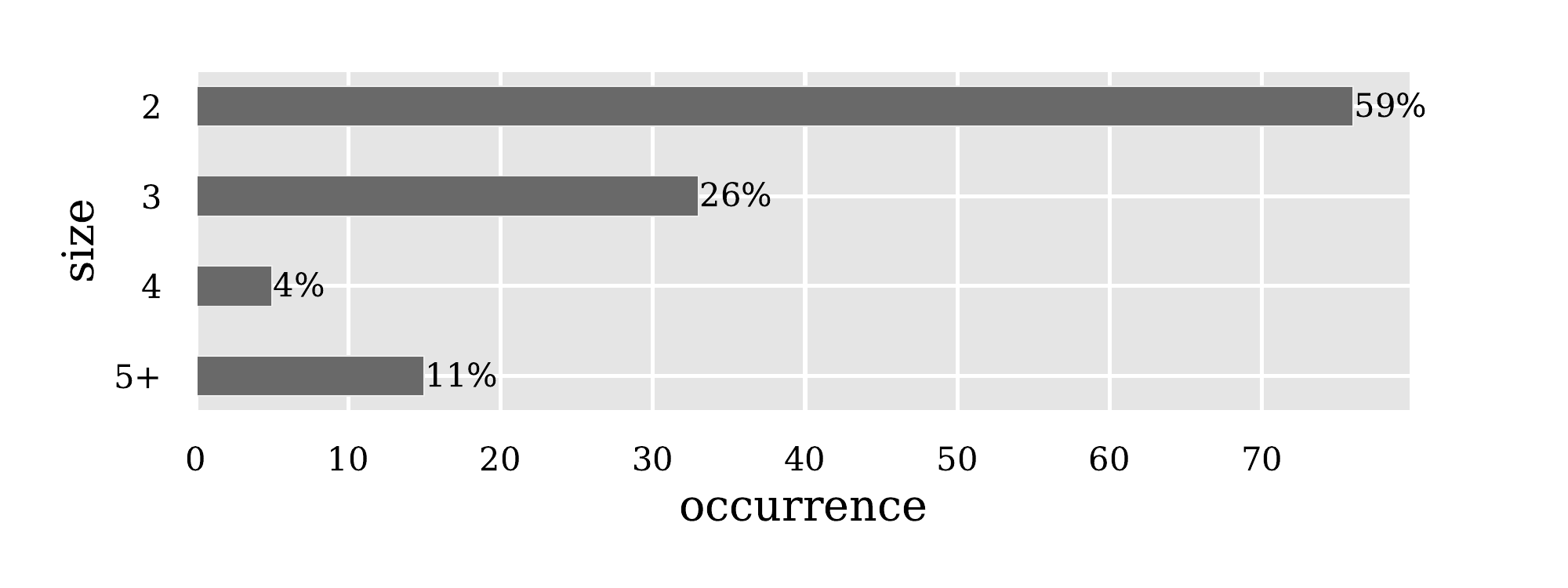}
	\caption{Size of composite refactorings (Composite Inline Method)}
	\label{fig:fig_degree_dec_inline}
\end{figure}

\bigskip

\noindent {\bf Age of composites.} In the oracle study (Section \ref{section:composite-oracle}),  we detect a single composite performed over multiple commits.  However, in the wild study, a significant part of the composites are performed over time (448 instances, 15.5\%). In these cases, we also assess age by computing the number of days between the most recent and the oldest commit in a composite.  Figure \ref{fig:fig_violin_composite_age_wild} shows the distribution of the results. As we can notice,  there are composites performed in a single day (3.3\%, 15 composites), but also there are composites performed over months.  Among the 448 composite instances, 75\% are performed up to 468 days (about 15 months), with a median age of 186 days (approximately six months). The 90th percentile is 835 days. However, it is difficult to generalize these results. For example, open source projects are subjected to multiple periods of inactivity~\cite{ist2020:GitHubMaintained}.

\begin{figure}[!th]
	\centering
    \includegraphics[width=.7\textwidth]{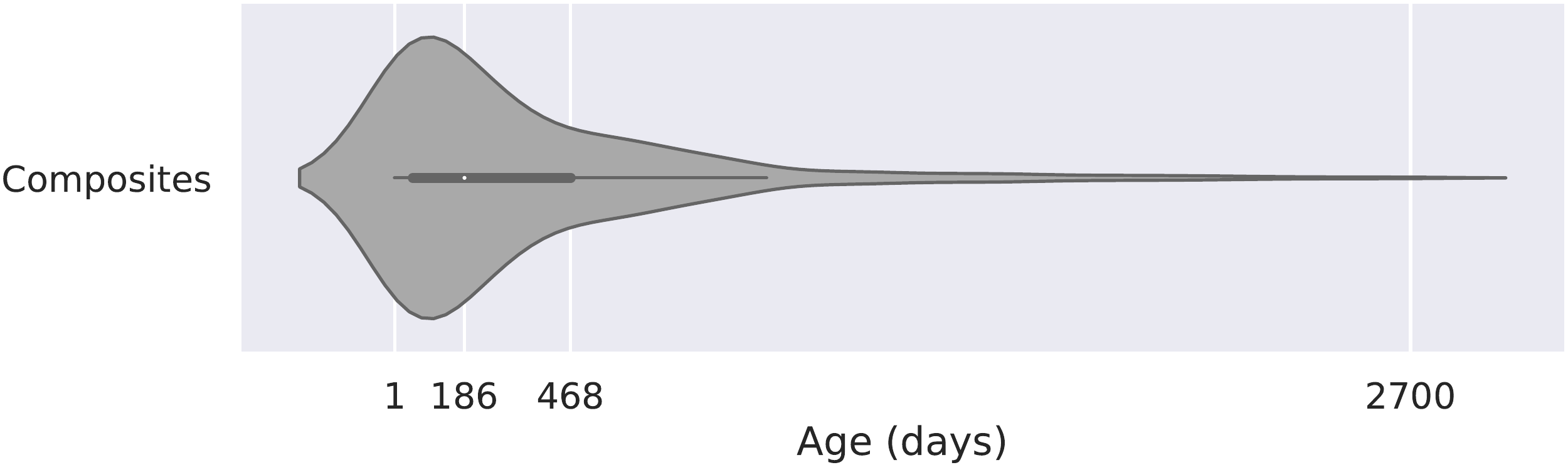}
	\caption{Distribution of the age of composite refactorings performed over multiple commits (wild, 448 instances)}
	\label{fig:fig_violin_composite_age_wild}
\end{figure}

\begin{tcolorbox}[left=0mm,right=0mm,boxrule=0.25mm,colback=gray!5!white]
\vspace{-0.2cm}
{\em Summary of RQ4:} 
In our extended dataset, most composite refactorings are also small, i.e., they are formed by two or three operations (2,258 composites, 78\%). 
Regarding the scope of the operations, most Method Decomposition are \textit{intra-class} (46\%), while most Method Composition are \textit{inter-class} (59\%). 
In other words, when decomposing methods, developers usually extract them to their current classes. In contrast, when removing code duplication, developers frequently extract methods from multiple classes.
\end{tcolorbox}

\noindent{\em Comparison with the oracle results (RQ2):}  In the study described in Section \ref{section:composite-oracle},  we also investigate composite characteristics. Regarding the size, the results are similar. For example, most composites have up to three operations (84\% in the oracle vs 78\% in the extended dataset).
The notable difference between both datasets refers to composites over multiple commits. The oracle only contains selected refactoring instances, i.e., it does not cover the whole projects' history. 
Due to this fact, we identified a single composite over time, i.e., a composite performed over more than one commit.  In contrast, in RQ4, we detect 448 composites spread over two or more commits (15.5\%). Finally, regarding the refactorings' scope, the results also follow a similar tendency. For example, most {\sffamily \small Method Decomposition} operations are {\em intra-class} in both samples. However, in the oracle, there is a higher frequency of {\em intra-class} operations (91\% in the oracle vs 46\% in the extended dataset).

\section{Discussion and Implications}
\label{section:discussion}

In this paper, we proposed a catalog of eight composite refactorings, i.e., refactoring operations composed of simple code transformations. Three of these refactorings---{\sffamily Class Decomposition}, {\sffamily Method Decomposition}, and {\sffamily Method Composition}---are new, in the sense they are not documented in Fowler's catalog. The other  refactorings---{\sffamily Pull Up Method}, {\sffamily Push Down Method}, {\sffamily Pull Up Field}, {\sffamily Push Down Field}, and {\sffamily Inline Method}---are also described in Fowler's catalog. However, we decided to include them in our catalog for two key reasons: (a) they imply the realization of multiple source code transformations that affect multiple program elements; (b) they are not properly detected by refactoring detection tools, such as RefactoringMiner~\cite{Tsantalis:2020:RefactoringMiner2} and RefDiff~\cite{danilo:tse2020:refdiff2}.
 However, to avoid potential conflicts, for these instances, we added the prefix ``Composite'' in their names. Regarding their popularity, the three new composites---{\sffamily Class Decomposition}, {\sffamily Method Decomposition}, and {\sffamily Method Composition}---represent about 77\% of the results in the wild study (Section  \ref{section:composite-wild}). In the oracle study, 88\% of the instances refer to these new cases (Section \ref{section:composite-oracle}). These values are highlighted in Table \ref{table:rq3-frequency}.

Essentially, the main contribution of our study is the catalog; the set of scripts to identify the described composite refactorings; and a new perspective of the well-known refactoring oracle proposed by Tsantalis and other researchers~\cite{TsantalisOracle,Tsantalis:2020:RefactoringMiner2}.

We claim this contribution can have two practical implications. First, as usual, our catalog highlights the importance and existence of composite refactorings. In other words, a catalog is a fundamental artifact to promote and disseminate the usage of composite refactorings among software practitioners. In fact, our studies showed that developers rely on composite refactorings during maintenance tasks. Therefore, the catalog and oracle can contribute to increasing the usage and application of such refactorings.

As a second practical implication, we showed that composite refactorings are not properly identified by refactoring detection tools, such as RefactoringMiner~\cite{Tsantalis:2020:RefactoringMiner2, Tsantalis:ICSE:2018:RefactoringMiner,pyref:scam:2021} and RefDiff~\cite{danilo:tse2020:refdiff2,danilo:msr2017:RefDiff,Brito:2020:RefDiff4Go}. Typically, these tools detect the parts of composite refactorings as independent operations. 
For this reason, we decided to implement a set of scripts to detect the eight composite refactorings in our catalog. 
Consequently, we also claim the concept of composite refactoring can be used to improve the results of  empirical software engineering studies on refactoring practices. 
Finally, our scripts and catalog can also help to improve the user experience provided by refactoring-aware code review tools~\cite{Brito:RAID:icpc:2021}, by supporting the detection of refactorings at a higher abstraction level.
}

\section{Threats to Validity}
\label{section:threatsValidity}

\noindent \textit{Generalization of results.} We characterized composite refactorings in terms of size and location. 
Our findings are based on a relevant oracle of refactoring operations and ten real-world Java systems hosted on GitHub. However, they---as common in empirical software engineering---cannot be generalized to other scenarios, such as closed software systems or other programming languages. 

\bigskip

\noindent \textit{Catalog of composite refactorings.} Our catalog includes eight composite refactorings, which describe a sequence of operations to compose or decompose source code elements.  
We acknowledge that the current version of our catalog is not complete and final. However, any catalog of refactorings can increase over time due to new insights, research, and development demands. For example, the first catalog proposed by Fowler has 68 refactoring operations~\cite{Fowler:1999}. After 18 years, in the second edition of his book, he introduced fifteen new refactorings~\cite{Fowler:2018}. 
We also followed the idea of Fowler's book~\cite{Fowler:1999,Fowler:2018}, using a single and popular programming language to guide the documentation and to provide illustrative examples. 
As mentioned by the author is ``\textit{better to use a single language so they can get used to a consistent form of expression}''.\footnote{\url{https://martinfowler.com/articles/refactoring-2nd-ed.html}} 
In fact, we plan to extend our study in the future, by including, for example, composite refactorings at the package level.
We also intend to explore other programming languages and refactoring types.

\bigskip

\noindent \textit{Detection of single refactorings.} 
Before detecting composite refactorings, we first need to identify single operations. In our first study---described in Section \ref{section:composite-oracle}---we rely on a well-known refactoring oracle, in which single refactoring instances were validated by multiple authors or tools~\cite{TsantalisOracle,Tsantalis:2020:RefactoringMiner2}. Therefore, our results are based on a trustworthy sample. 
For the second study---described in Section \ref{section:composite-wild}---we rely on RefDiff~\cite{danilo:tse2020:refdiff2} to mine refactorings in ten popular projects.
According to recent results, the precision of RefDiff is high, reaching 96.4\% for Java~\cite{danilo:tse2020:refdiff2}.
In this second study, we also cleaned up the dataset, for example, we remove packages that are not part of the core system (e.g., \textit{test}, \textit{docs}, \textit{sample}), and we removed constructors since they are essentially initialization structures.
Finally, as natural during software evolution,  commits can include temporary or unintentional operations, such as reverted commits due to test fails and experimental code. To mitigate this threat, we focus only on the main branch evolution.

\bigskip

\noindent \textit{Detection of composite refactorings.} 
Regarding composite detection, we implement a set of scripts, as described in  Section \ref{sec:scripts-detect-composite}. 
The input comprises a list of single refactoring operations, including details such as path, refactoring type, and entities names. 
A possible threat is the possibility
of errors in the implementation of our tool and parsers.
For the oracle analysis, we extract this information from textual data. We also rely on well-known Python libraries to mitigate this threat, e.g., retrieving the data by regex expression. 
Also, we inspected a sample of 28 composite refactorings to check the results (see details  in Section \ref{sec:scripts-detect-composite}), when we did not identify any error in the process of clustering refactoring operations as composites. Our verification included 160 single refactoring operations from the oracle created and curated by Tsantalis et al.~\cite{Tsantalis:2020:RefactoringMiner2,TsantalisOracle,Tsantalis:ICSE:2018:RefactoringMiner}. Finally, we are making publicly available the datasets and scripts used to detect composite refactorings.
\\[-0.2cm]

\section{Related Work}
\label{section:relatedWork}

We organized related work in three subsections: (a) field studies regarding sets of related refactoring operations; (b) studies about catalogs; and (c) other studies on refactoring.

\subsection{Batch and Composite Refactorings}

Refactoring was already studied in scenarios such as code review~\cite{hayashi2013rediffs, ge2014towards,ge2017refactoring,Brito:RAID:icpc:2021}, code understanding~\cite{danilo:fse2016:WhyWeRefactor,Wang:ICSM:2009,Pantiuchina:2020,Yaroslav:FSE:2021}, and education~\cite{Lopez:2014:SIIE,Hebig:2020:ICPC}. However, these studies do not propose catalogs of refactorings operations to improve software practices. Most of them also focus on single refactoring operations.

There are two central types of studies regarding groups of related refactoring operations,  studies on {\em batch} refactorings and studies on {\em composite} refactorings, which is the concept we explored in this paper. Batch refactorings refer to a set of single refactoring operations, which are then grouped considering criteria such as time~\cite{Murphy-Hill:ICSE:2009,Murphy:2012:TSE}, version system~\cite{cedrimRego:2018}, and developers~\cite{Bibiano:esem2019:BatchRefactoring,cedrimRego:2018}. As mentioned by Cedrim \textit{et. al.}~\cite{cedrimRego:2018},{\em ``the way the batches are synthesized is open-ended, i.e., different developers can have different views of how to create a batch''}. Similarly, composites are defined as sequences of atomic refactoring operations~\cite{Sousa:2020:Composite,FASE:DSLErlang:Composite:2012,Tsantalis:2020:RefactoringMiner2}. This concept is explored in contexts like domain specific languages for describing refactoring~\cite{FASE:DSLErlang:Composite:2012} and code smells~\cite{Sousa:2020:Composite,Bibiano:ICSME:2021}.

Sousa \textit{et. al.}~\cite{Sousa:2020:Composite} originally defined composite refactorings as {\em ``two or more interrelated refactorings that affect one or more elements''}. 
The detection of composite refactoring relies on three distinct heuristics.
The first heuristic--called \textit{element-based}--combines single refactoring operations by the scope.
The scope can be, for instance, a single class. 
For example, the authors show a composite refactoring from this category, which includes the movement of attributes, movement of methods, and extract superclass operations.
The \textit{commit-based}  heuristic links refactoring operations performed in a single commit.
Finally, the third heuristic--named \textit{range-based}--connects refactorings by location (e.g., if a refactoring crosscuts two classes named $C_1$ and $C_2$, both are part of the location). As a consequence, an instance of a composite can include mixed operations at distinct levels, i.e, classes, attributes, and methods.
In summary, the study considers some criteria to cluster composites, also reusing previous heuristics~\cite{Bibiano:esem2019:BatchRefactoring,cedrimRego:2018}. However, they do not introduce and document a catalog of composite refactorings (as we do in this paper) and a significant part of the study  investigates the relevance of composites for removing code smells.  Although we are reusing the definition, in this paper, we explore another perspective, i.e., our key goal is to propose and document a catalog of composite refactorings. Moreover, we also show the importance of composite refactorings by mining and characterizing their occurrence in two datasets: a sample with hundreds of confirmed single refactoring operations and the history of ten well-known open-source projects.

There are also studies focusing on subcategories of composite refactorings. 
For example, ``incomplete composites''~\cite{Bibiano:Incomplete:2020}, i.e.,  when the composite refactoring \textit{``is not able to entirely remove a smelly structure''}.

\subsection{Catalog of Refactorings}

Recently, Bibiano et. al.~\cite{Bibiano:ICSME:2021} investigated “complete composites”, i.e., sets of refactoring operations that remove the whole occurrence of four code smell types. The study includes 618 complete composites formed by well-known refactoring operations. Differently from our study, the identification of composite refactorings relies on a range-based heuristic defined in previous studies~\cite{Sousa:2020:Composite},  which groups refactorings affecting the same location. The authors also present a catalog of complete composites to remove code smells. This catalog includes five complete composites,  which are sequences of {\sffamily Move Method} or {\sffamily Extract Method} operations. For example, their catalog focuses on the removal of Long Method (i.e., large and complex methods) and Feature Envy (i.e., a method that uses several methods from a distinct class).  For each instance, the authors discuss side-effects, i.e., when a composite removes a target code smell but introduces other ones. Among the five complete composites from their catalog, three instances refer to extract operations to remove long methods. However, in the first case, the extraction contributes to introducing a Feature Envy. The second one does not reduce the method’s size, i.e., it is necessary to perform new extract operations to remove the smell. Finally, the third instance introduces a long parameter list before the extraction.  In our study, we propose a catalog of eight types of composite refactoring, which are formed by distinct refactoring types. We cluster refactoring operations by considering the source or target code elements. In other words, our scripts identify sequences of single refactoring operations to compose or decompose a source code element, regardless of the presence of a code smell. In fact, there are several reasons to refactor a given source code element, which do not necessarily involve smell removal~\cite{danilo:fse2016:WhyWeRefactor,Pantiuchina:2020}.

Tsantalis et. al.~\cite{Tsantalis:2020:RefactoringMiner2} also present a brief discussion regards composites. The authors introduce a new version of RefactoringMiner, which detects {\small \sffamily Extract Class}---also defined in Fowler's catalog~\cite{Fowler:2018,Fowler:1999}. 
According to the authors, composite refactorings {\em ``are composed of basic ones''}.  
Therefore, {\small \sffamily Extract Class} matches  this concept, since it comprises a set of {\small \sffamily Move Method} and {\small \sffamily Move Field} operations aiming to generate a new class. 
In our catalog, {\small \sffamily Class Decomposition} can include a set of move operations to a new class or existing one. However, the focus refers to the decomposition of the source class.

Fowler proposes a popular and widely used catalog of refactoring operations~\cite{Fowler:2018,Fowler:1999}. 
The recent version includes new composite refactorings, such as {\small \sffamily Inline Class} and {\small \sffamily Collapse Hierarchy}.\footnote{\url{https://refactoring.com/catalog}}
In the case of {\small \sffamily Inline Class}, we eliminate a class by moving all elements to distinct ones. 
Therefore, it is a subcategory of {\small \sffamily Class Decomposition}. 
However, the current refactoring detection tools do not support this composite~\cite{Tsantalis:2020:RefactoringMiner2,Tsantalis:ICSE:2018:RefactoringMiner,danilo:msr2017:RefDiff,danilo:tse2020:refdiff2,Brito:2020:RefDiff4Go,pyref:scam:2021}.
In {\small \sffamily Collapse Hierarchy}, we eliminate subclasses by moving all elements to the superclass. Therefore, {\small \sffamily Composite Pull Up Method} can be a part of this operation. The current version of RefactoringMiner detects this refactoring~\cite{Tsantalis:2020:RefactoringMiner2}. However, it is not properly explored in the literature. For example, the oracle used in this paper includes only a single instance of {\small \sffamily Collapse Hierarchy}.

\subsection{Other Studies on Refactoring}

In previous papers, we explored the reasons for refactorings performed over time~\cite{Brito:2021:RefGraph:EMSE,Brito:2020:RefGraph}. We analyzed characteristics such as time, refactoring types, and authorship. In particular, we relied on a graph-based abstraction---called refactoring graph---to mine refactoring operations performed over the history of ten GitHub projects.  The insights from this study helped us to propose  the catalog of composite refactorings described in this current paper.

Finally, we reinforce findings from a recent study that points out a significant rate of multiple extractions to decompose methods in a single commit~\cite{Hora:emse:2020}. The authors show that Extract Method operations are frequently performed by developers, who create methods for distinct purposes, such as testing, validation, and setup. However, the study does not document a catalog of composite refactorings. Its goal is to characterize method extractions, for example, their content, size, and degree. Similarly, in our study, Method Decomposition and Method Composition—composites formed by Extract Method and Extract and Move Method operations—are among the top-3 most frequent composites. In the study in the wild, for example, we detected 1,265 occurrences. Among them, 275 composites (21.7\%) involving extractions are performed over multiple commits.

\section{Conclusion}
\label{section:conclusion}

We introduce a catalog of composite refactorings.
According to our definition, a composite refactoring can be spread in multiple commits.
Our catalog includes eight instances that describe sequences of  operations that compose or decompose program elements: {\sffamily Method Composition}, {\sffamily Method Decomposition}, {\sffamily Class Decomposition}, {\sffamily Composite Pull Up Method}, {\sffamily Composite Push Down Method}, {\sffamily Composite Pull Up Field}, {\sffamily Composite Push Down Field} and {\sffamily Composite Inline Method}.

In order to show that the proposed refactorings occur in real scenarios, we searched for occurrences of each instance in two datasets. First, we focus on a well-known refactoring oracle. In this first study, we identify that about 60\% of the selected sample is part of a higher-level composite refactoring. Then, we mine the history of ten popular GitHub projects, in which we detected 2,886 instances of composite refactorings.
 
Future work might include an extension of the proposed catalog with other composite refactorings.
 We also plan to extend our mining study with projects implemented in other programming languages, such as C, JavaScript, and Go. 
 The scripts are publicly available at  \url{https://github.com/alinebrito/composite-refactoring-catalog}

\section*{Acknowledgments}

\noindent This research is supported by grants from FAPEMIG, CNPq, and CAPES.


\bibliographystyle{plain}
\bibliography{bibfile}%

\clearpage

\section*{Author Biography}

\parpic{\includegraphics[width=39mm,clip,keepaspectratio]{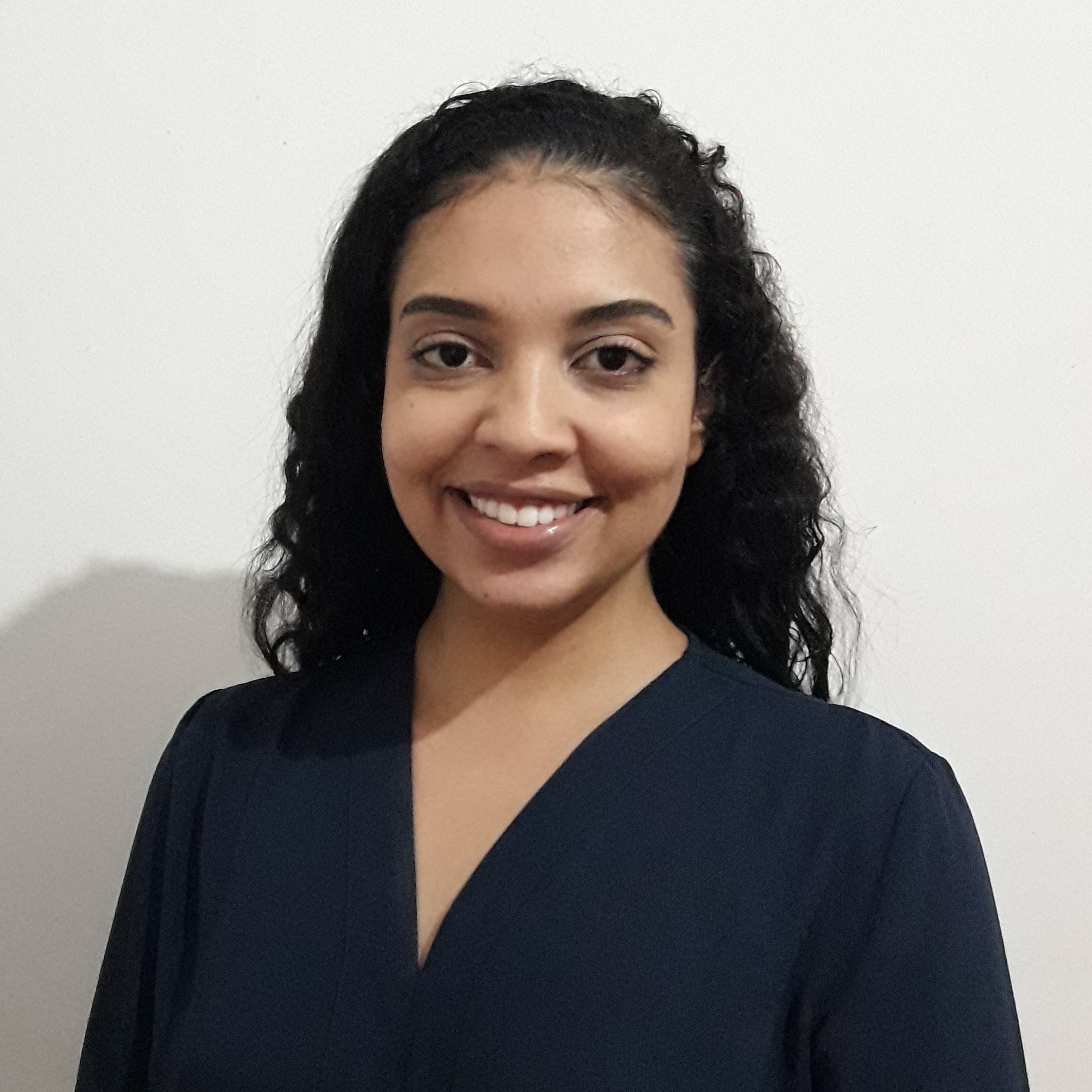}} \sloppy 
\noindent {\bf \small Aline Brito} \small is a PhD candidate in the Computer Science Department at the Federal University of Minas Gerais (UFMG), where she also received a Master's Degree in Computer Science. She received a Bachelor's Degree in Computer Engineering from the Pontifical Catholic University of Minas Gerais (PUC Minas). Brito also was a software developer for  five years. Her research interests include software quality analysis, software maintenance and evolution, and software repository mining. Contact her at \url{alinebrito@dcc.ufmg.br}; \url{alinebrito.com}.\vspace{0.7cm}

\parpic{\includegraphics[width=39mm,clip,keepaspectratio]{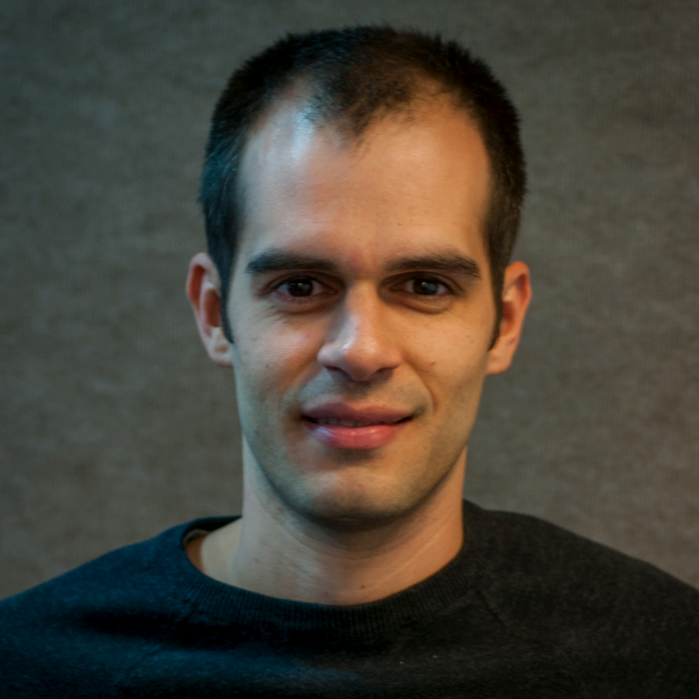}}
\noindent {\bf \small Andre Hora} \small  is a professor in the Computer Science Department at the Federal University of Minas Gerais (UFMG). His research interests include software evolution, software repository mining, and empirical software engineering. Hora received a PhD in Computer Science from the University of Lille. He was a Postdoctoral researcher at the ASERG/UFMG group during two years and a software developer at Inria/Lille during one year. Contact him at \url{andrehora@dcc.ufmg.br}; \url{www.dcc.ufmg.br/~andrehora}.\vspace{1cm}

\parpic{\includegraphics[width=39mm,clip,keepaspectratio]{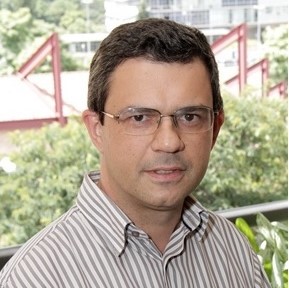}}
\noindent {\bf \small Marco Tulio Valente} \small is an associate professor in the Computer Science Department at the Federal University of Minas Gerais (UFMG), where he also heads the Applied Software Engineering Research Group (ASERG). His research interests include software architecture and modularity, software maintenance and evolution, and software quality analysis. Valente received a PhD in Computer Science from the Federal University of Minas Gerais. He is a Researcher I-D of the Brazilian National Research Council (CNPq) and holds a Researcher from Minas Gerais State scholarship, from FAPEMIG. Contact him at \url{mtov@dcc.ufmg.br}; \url{www.dcc .ufmg.br/~mtov}.\vspace{0.1cm}

\end{document}